# THE EVOLUTIONARY BIOLOGY OF OURSELVES: UNIT REQUIREMENTS AND ORGANIZATIONAL CHANGE IN UNITED STATES HISTORY

Carl P. Lipo
Department of Anthropology
Box 353100
University of Washington
Seattle, WA 98195-3100
(206) 543-5240
(206) 543-3285 FAX
clipo@u.washington.edu

Mark E. Madsen
Emergent Media, Inc.
411 Fairview Ave. N.
Suite 200
Seattle, WA 98109
Madsen@emergentmedia.com

## INTRODUCTION

Concerning the origins of "modern" American society from simpler beginnings, historian Henry Steele Commager wrote:

> "In the generation after Appomattox the pattern of our present society and economy took shape. Growth--in area, numbers, wealth, power, social complexity, and economic maturity--was the one most arresting fact. The political divisions of the republic were drawn in their final form … and an American empire was established" (Nevins and Commager 1956:241).

Theories of cultural evolution in their various forms posit that cultural development occurs as a consequence of the expansion of societies coupled with increased control over social and technical activities (e.g., White 1949; cf. Harris 1968). Growth necessarily precedes organizational change because it provides the impetus for change. Studies of complex societies, whether prehistoric, historic, or contemporary have frequently taken this stance.

Numerous challenges have been made to cultural evolution. Over the past decade scientific, or Darwinian, evolutionary theory has become an increasing focus of anthropological research into cultural change. Interest in Darwinian theory is evident both in general anthropology (e.g., Boyd and Richerson 1985; Rogers 1995; Smith and Winterhalder 1992) and in archaeology (e.g., Barton and Clark 1997; Dunnell 1980, 1989, 1992, 1995; Feathers 1990, 1993; Leonard and Jones 1987; Leonard and Reed 1993; Lipo et al. 1997; Lyman and O'Brien 1998; Maschner 1996; Neiman 1993, 1995; 1997; O'Brien 1996; O'Brien and Holland 1990; Rindos 1984, 1989;Teltser 1995). Crucial to anthropological interest in Darwinian evolution is expansion of the general Darwinian paradigm to include models of cultural transmission, a step which has yielded important conceptual models and methodological insights (Boyd and Richerson 1985; Cavalli-Sforza and Feldman 1981; Dunnell 1989; Lipo et al. 1997; Neiman 1995). Theory development, however, has rarely been concerned with the specifics of connecting models of general process to the empirical subject matter of particular fields, in this case archaeology. While formalized descriptions of Darwinian processes are excellent starting points for theory development, the success of theory hinges not upon the quality of quantitative models, but upon the explanation of empirical cases. Explanation requires that quantities required by models be empirically measurable, using units that are specified by theory as meaningful (Lewontin 1974; Dunnell 1986).

Units are concepts where generalized processes (such as natural selection) interface with the subject matter and empirical limitations of a particular discipline, such as biology, archaeology or economics. Therefore, when theories omit construction of units, application of models to empirical cases becomes difficult. Theoretical models are more than simply descriptions of processes. Models are a fusion of units and relationships between units: the relationships between units in a model define a process while units specify the scales of entities upon which processes operate and how the measurements required by the model are made. Despite the fact that models are often stated in the form of mathematical equations, units are not simply variables in equations. Units specify how quantities required by mathematical equations (e.g., frequencies) are measured in empirical situations in terms that are meaningful. Exclusion of units from consideration in the theory development process has two critical results: explanation of real-world cases becomes impossible and further model development is inhibited insofar as alternative formulations cannot be disqualified on empirical grounds.

The study of the evolution of complex societies is an excellent case in point. The developmental relation between "simple" and "complex" societies is a major transition in human history and has been the subject of extensive investigation and speculation (e.g., Adams 1966; Athens 1977; Braidwood 1937, 1962; Childe 1925, 1941, 1958; Flannery 1972; Friedman 1974; Service 1975; Spencer 1883; Tylor 1865; Wright 1977, 1984, 1986). One feature, common to all views of complex society, is that there is a quantum increase in the scale of organization and the concomitant appearance of specialization, especially craft specialization, that accompanies the emergence of complex societies. Anthropologists have used a wide variety of theoretical stances to explain this phenomenon including cultural evolution (e.g., Carneiro 1970; Sahlins and Service 1960; Segraves 1974; Service and Cohen 1978; Steward 1949, 1968; Tylor 1865; White 1949), cultural ecology (e.g., Boserup 1965; Clarke 1968; Flannery 1972; Rappaport 1968; Redman 1978; Wittfogel 1957), and Marxist structuralism (Bloch 1975; Diakonov 1969; Friedman 1974; Friedman and Rowlands 1977; Spriggs 1984). In general, these approaches to the study of the archaeological record share the notion that social complexity results as a cultural response to population growth and pressure. Despite over a century of research, however, these perspectives have produced few satisfactory explanations and no general theory for explaining why complex societies evolve when and where they do.

In an effort to integrate evolutionary theory into the study of complex societies, Dunnell and Wenke (1980; Dunnell 1978) proposed that in evolutionary terms, the distinction between simple and complex societies is of fundamental importance and can be expressed as the scale at which selection operates on human populations or the scale of the evolutionary "individual" (Dunnell 1978; 1995). In their framework, simple societies are exist when the scale of selection is that of the organism, i.e., persons (ignoring age and sex) carry the full codes, genetic and cultural, to reproduce the human phenotype. Simple societies are aggregates of such redundant units. Many other, less archaeologically-visible features of simple societies (e.g., kin-based organization, intrasocietal competition, etc.) are thus explained as well (Dunnell 1978, 1980, 1995; Dunnell and Wenke 1980). Complex societies, on the other hand, are populations in which the scale of selection, the evolutionary individual, is larger than the organism, i.e., persons do not carry the entire set of instructions for producing a human phenotype. Societies, or units of similar scale, are the individuals upon which selection acts. In complex societies, it is only at the scale of aggregates of organisms at which all of the functional requirements for reproduction of the society are met. Within complex societies, people perform different functions, each of which is necessary to the survival of the whole. Each subpart performs only one of a few functions out of the total --- subparts cannot survive on their own. This element is recognized archaeologically by such things as craft specialization, polity-based organization, wide scale functional interaction and inter-societal competition (Dunnell 1978, 1980, 1995; Dunnell and Wenke 1980).

In spite of its potential, Dunnell and Wenke's evolutionary approach to the study of complex societies has seen little use in archaeological research. The lack of widespread acceptance is primarily due to three reasons. First, critics of this evolutionary approach usually argue that the mechanism for change in the Dunnell and Wenke model is group selection and that, following common wisdom, group selection should rarely, if ever, be invoked in evolutionary explanations. However, as Sober and Wilson (1998; see also Wilson and Sober 1994; Wilson 1997) recently report, there is every reason to suspect that a multilevel selection argument is appropriate for explaining group level phenomena. Second, empirical evaluation is hampered by the lack of measuring units by which one can go out and examine the world archaeological record. Beyond some vague notions, we have no grammar for building descriptions of the functional organizational change that is concomitant with evolutionary theory. Finally, empirical evaluation of this framework in the archaeological record is extremely difficult due to the lack of data that is of the appropriate scale and precision.

Our goal, therefore, is to construct measurement units by which it will be possible to measure organizational change within an evolutionary framework. After summarizing the most widely agreed upon criteria of organization, we suggest means of creating units for models of selection upon culturally transmitted variation that do not rely upon assumptions of traditional units or pre-defined scales of analysis. In order to demonstrate how these units are useful for examining the evolution of functional organization in the empirical world, we turn to a historical example where there exist abundant and readily available data for this task. We demonstrate that at least some of the functional changes involved in shifting scales of selection occur in the history of the United States economy, from the middle of the nineteenth century through today, the period over which Nevins and Commager (1956) argue that essentially "modern" features of contemporary American society developed.

## THE IMPORTANCE OF UNITS

Units are central for the construction of empirically useful models. Because of their centrality, it is perhaps surprising that existing formalizations of Darwinian processes in anthropology do little to address unit requirements. The long experience of biologists, though, is instructive. Theories are rarely created all at once and in finished form. Lewontin (1974) outlined a cycle of steps for the building and empirical evaluation of biological theories. He noted that theory-building is an iterative process, proceeding from descriptions of process with ad hoc variables to the creation of units used to examine the model's performance in explaining particular empirical problems. Rather than a linear process, Lewontin stressed that the process was a halting, back-and-forth enterprise, typically requiring many iterations through revision of units and descriptions of process before arriving at a robust model with empirical utility (Figure 1). The current situation in anthropological attempts to apply Darwinian models to cultural phenomena reflects only the youth of our efforts, not fundamental problems with the application of evolutionary theory to humans.

Continued development of models without formalized units creates a danger. Occasional attempts at using incomplete models to explain empirical cases often produce intuitively satisfying conclusions, which Gould and Lewontin (1979) label “just-so stories. Even if models are rigorously stated in formal, mathematical terms, without empirically measurable units the articulation of the model with the real world is cast entirely in terms of units derived from ethnocentric common sense or disciplinary tradition. Both anthropology and evolutionary biology are saddled with such units (Hull 1980, 1988; Dunnell 1989, 1992), to the extent that each discipline has not reanalyzed the way in which empirical units are defined and used (Table 1).

As long as units and unit requirements for particular models remain unanalyzed, we will be able to falsify hypotheses about the utility of a “model” in a particular empirical case only fortuitously. This can occur when the commonsensical units we employ actually come close to meeting the largely unspecified requirements of the incomplete models we use for substantive research. A better way to proceed is to work on both sides of Lewontin's cycle at once: writing better process models and considering the unit requirements of those models for each of the sub-field's particular data sources simultaneously.

Because discussion of evolutionary process has outpaced discussion of unit requirements and development of units, the purpose of this chapter is to consider in detail the unit requirements for theories of selection upon culturally transmitted variation. Theories using selection will be the most common models used to explain many obvious features of human behavior and the archaeological record (e.g., subsistence and settlement), based upon the supposition that such features greatly affect the fitness of individuals. In biology there is considerable discussion about the requirements for an entity to be involved in a selection process, and especially over the set of empirical entities at different scales that meet these requirements. Of particular interest are the conditions under which sets of organisms can act as a single unit of selection, given Dunnell’s (1978; Dunnell and Wenke 1980) model of the origin of complex societies as a shift in the scale at which selection acts.

## UNITS AND UNIT REQUIREMENTS FOR SELECTION MODELS

Within evolutionary biology and more recently, within the philosophy of biology, there has been a great deal of controversy over the empirical entities upon which selection acts. There has also been discussion, though subordinate in the literature, concerning the defining criteria for units of selection (e.g., Dawkins 1982; Hull 1980, 1988; Kawata 1987; Lewontin 1970; Sober 1984, 1992, 1993; Williams 1966, 1992). The first debate is an empirical one and we need not wait for any resolution before proceeding with anthropological unit formation and model building. The second discussion is more useful for our purposes as the resultant literature clarifies the criteria particular entities must meet to be units upon which selection acts. It is this second endeavor we are interested in extending to the human case. There are two steps to this process. First, we outline definitions of necessary units, and second, we refine these definitions to the point where methods for measurement can be constructed. Limitations of space prevent us from taking up specific methods except in broad outline.

### Kinds of Units Required in Selection Models

#### Theoretical and Empirical Units

Two kinds of units are necessary to link theory and models with empirical cases (Dunnell 1986). Theoretical units are measurement scales that are specified in the definition of a process; they denote sets of empirical entities that are used in theory evaluation or falsification. In order for a theory to function in explanations of the real world, processes must be defined that include the appropriate units so that confrontation of the theory by empirical cases is meaningful (dynamic sufficiency, *sensu* Lewontin 1974). In addition to dynamic sufficiency, units defined in theory must be measurable in the real world to allow explanations to be falsified and theory evaluated for its utility (empirical sufficiency, *sensu* Lewontin 1974). The second class of units is empirically defined. Empirical units are generated by using theoretical units to measure the real world and group sets of entities.

Although the distinction between theoretical and empirical units is perhaps elementary, its importance cannot be overstressed. Inattention to this distinction is what leads some to write theoretical models with common-sense entities as units, and many controversies in biology can be traced to this source. Confusing the roles of theoretical and empirical units is what generates the controversy over "group" versus "individual" selection, for example. Hull (1980: 311-332) points out that when biologists address the question of the units of selection, they take the traditional hierarchy of organization as fundamental and the level at which selection operates as variable:

> "Evolutionary biologists are currently confronted by a … dilemma: If they insist on formulating evolutionary theory in terms of commonsense entities, the resulting laws are likely to remain extremely variable and complicated; if they want simple laws, equally applicable to all entities of a particular sort, they must abandon their traditional ontology. "

Two entities that each meet the criteria for functioning within an evolutionary process must be considered as the same kind of entity even if one happens to be an organism and the other a colony of functionally interdependent organisms. Thus, Hull argues, there is no such thing as “group” or “individual” selection, merely selection that operates upon entities that can occur at multiple scales at once. In order to avoid this confusion and to make advances in understanding the evolution less well understood phenomena (such as the evolution of sets of culturally transmitting entities), it is necessary to develop a model that establishes the necessary units and define units that measure quantities meaningful in the model (Figure 2).

Hull’s approach opens the interesting possibility of mapping the changing scales at which selection is operative in particular environments or over specific spans of time as significant phenomena requiring evolutionary explanation in their own right. For example, in a *selection* model phrased in terms of organisms, as was Darwin’s original model and the “evolutionary synthesis”, it is impossible to account for the origin of “organisms” as a distinctive organizational phenomenon *using natural selection* as the causal force. The units we have chosen to use, in this example, force a transformational rather than selectionist explanation of the origin of “organisms” since in model there is no variability — selection acts upon “organisms,” period. If a model assumes the presence of a specific scale or kind of entity, then we logically cannot use the model to account for the origins of that particular kind of entity (Buss 1987).[1] In a selection model phrased in generalized theoretical terms, however, variation in scale of organization becomes immediately explicable by selection, since particular kinds of organization are not longer *assumed* by the model.

With this in mind, we can appreciate that many of the traditional debates are simply not problematic — despite claims to the contrary, genes are not better than organisms as units of selection (Buss 1987; Sober 1984; cf. Dawkins 1976, 1982; Williams 1966, 1992). Similarly, there is no reason why groups of organisms that meet the criteria for a unit of selection cannot evolve via natural selection. Resolution of the controversy is an empirical question soluble only in different times, places and *especially* in the context of different research questions. Selection will act upon any entity that meets a set of defining criteria (Lewontin 1970). If single clumps of genetic material, functionally integrated organisms, and colonies of organisms variously meet these criteria, then we are learning something interesting about the evolution of life, and should not regard multiple scales of selection as presenting a matter for disagreement. As Sober and Wilson (1998) discuss in their recent book, the empirical controversy over the reality of multi-level selection is largely over in serious biological circles; only the philosophical argument remains. Discussion of "group" selection within evolutionary anthropology, however, has largely followed the standard critique and "individual" selection accepted as the only legitimate focus of research (e.g., Boone and Smith 1998; Neiman 1997; Smith 1991; Smith and Winterhalder 1992). We believe that it is time to follow Hull's lead and examine the utility of multi-level selection in the cultural case, particularly with respect to so-called "complex" societies.

## <u>Unit Requirements and Scale</u>

In addition to different kinds of units for theoretical definitions and empirical entities, natural selection models require theoretical and empirical units at multiple scales (Figure 3). Since evolution and selection are measured as changes in the frequencies of individuals sharing particular attributes within a well-defined population (Lewontin 1974; Sober 1984, 1992, 1993), three kinds of units are necessary. First, we need to define a unit in which entities are the focus of selection forces. This unit represents the entities that interact with their environment in such a way as to bias probabilities of survival and reproduction. This unit is generically termed the *interactor* by Hull (1980) and Sober (1984, 1993)[2] and we follow this usage here. These theoretical units can occur at any number of actual scales in the empirical world. In general, two classes of entities meet the definition for interactors. The first we refer to as *individuals,* which in the biological literature refer to spatio-temporally bounded and continuous entities (e.g., Hull 1980). Kawata (1987) notes that his definition for "integrated units" (of which the interactor is an example) and "individual" are not equivalent, and it is easy to appreciate why. Interactors are theoretical units, and individuals are empirical entities, some of which meet the criteria for being interactors and some which do not. The second class of entities is what D.S. Wilson (Wilson 1975; Sober and Wilson 1998) refers to as "trait groups." Unlike individuals, trait groups are *sets* of entities that are defined on the basis of traits that they share through interactions without requiring discrete spatial or temporal boundaries. As Sober and Wilson (1998) have shown, trait groups are necessary for examining the evolution of traits in populations where it may be disadvantageous for any one individual to have that trait but good for the trait group to share. This process may, in fact, explain the evolution of individuality (Buss 1987) but for the present purpose of understanding functional organization and scale, our discussion is restricted to individuals and their definitions.

---

[1] The example and the point are derived from Dawkins (1982), which contains an elegant discussion of “organism-ness” as a contingency-bound phenomenon requiring the same kind of attention to historical explanation that other phenomena routinely attract. Our point is more general, however, in that empirical entities should not be written into models as assumptions or as substitutes for theoretical units. Rigorous attention to the distinction between theoretical concepts on the one hand and empirical entities on the other is required at every step in order to prevent models from including cryptic commonsensical inputs that will ultimately foil attempts to use the model.

The second scale of units required by selection models is smaller than the interactors; these will be *attributes* or characteristics of interactors. Interactors share the defining criteria for being units of selection but are variable with respect to units at smaller scales, and it is this variability that gives rise to the force of natural selection. Research problems define relevant classes of attributes by reference to theories that address the physical and ecological basis for interaction between interactors and their external environment (Dunnell 1995; Sober 1984). Derived from theory, attribute classes are entirely conceptual and thus can vary in empirical scale as the interactor varies in empirical scale. The entities that correspond to attribute classes are the actual phenotypic attributes of interactors, abbreviated here to *phenotype* in accord with conventional usage (Dobzhansky 1970) of the term to mean the outward behavior, body, and manufactures of an organism.

A third unit is required at a scale larger than the interactor, and comprises aggregates within which we will measure change (as differences in frequency through time of interactors that share attribute classes). To distinguish it from *populations* of empirical entities we simply refer to the theoretical unit as the *aggregate*. This unit is similar in some ways to the common-sense idea of population, differing because aggregates are defined with respect to a particular research problem and are entirely conceptual (Sober and Wilson 1998). For example, if we are interested in explaining change in the sex ratio of interactors, the proper aggregate would probably correspond to the biological unit called "deme," or interactive breeding group (Maynard Smith 1989). If, however, we are interested in explaining why a particular mobility pattern is increasing in frequency within a region, the proper aggregate will be defined according to the physical and biotic environment. Choice of the improper aggregate for a given problem can easily foil attempts at creating empirically usable models, by creating populations for sampling that are irrelevant to the problem of interest, as a simple example demonstrates (Figure 4).

Selection models are written in terms of relative frequencies rather than absolute abundances or counts (Lewontin 1974), and frequencies depend strongly upon the choice of counting frame within which to derive proportions. If, as in Figure 4, we are interested in the evolution of sex ratio for a set of interactors, we want to sample interactors that belong to the same breeding population in order to calculate frequencies. The solid lines in the figure represent actually breeding groups of interactors. If we define our aggregate using a commonsensical notion of population, such as spatial continuity, we might be lead to delineate populations indicated by the dotted lines. The frequencies of sexes for the breeding populations (solid lines) and sampled populations (dotted lines) demonstrate the problem. A population that includes individuals from more than one breeding group yields frequencies characteristic of neither breeding group. The population sampled from a single breeding population yields frequencies nearly identical to the frequencies of the larger breeding group, the desired result.[3] The general point here is that, for a specific problem, the aggregate of interactors must be defined such that a group of entities is subject to a common selective environment (Sober 1984), and populations must be delineated in the real world that use the aggregate definition to bound a group of interactors (Sober and Wilson 1998). Only in this way will frequencies be meaningful to evaluation of our models and explanation of empirical cases.

---

[2] In the literature concerning the unit requirements for selection models, there are a large number of terms that refer to the same kind of unit; namely, the unit which varies in fitness and consequently creates selection by differential survival or replication. We have chosen the term *interactor* (Hull 1980; Sober 1984, 1993) because it captures the essence of the entity/environment interaction that causes selection. Other terms include vehicle (Dawkins 1976, 1982), and generically throughout the literature, "unit of selection." The term "level of selection" is often used to refer to the scale of entities at which selection appears to be occurring but without specifying units or entities explicitly.

[3] Assuming, of course, that the sample was properly drawn from the statistical population, that sample sizes are large enough to accurately estimate the population mean, and so on.

Another way to look at units and scale is to use Sober's (1984) concepts of *selection of* and *selection for*. Sober points out that while selection operates upon interactors, creating differential survival and replication, selection operates *in consequence* of variation in the attributes of interactors. While there is selection *of* entities, this always occurs via selection *for* properties. This distinction is useful because it not only emphasizes the role of the two smaller scales at which units need to be constructed, but it indicates the starting point for unit construction and theory-building. Since selection is observed as differences over time in the frequencies of interactors sharing particular classes of attributes, it is typically intuitive perceptions of attribute changes that generates the search for an evolutionary explanation. We often know, approximately, some of the relevant attributes before we know the relevant interactors or aggregates.

Research problems and the conditions of specific situations drive the business of defining interactors (Kawata 1987:416). Characteristically, analyses for scales of selection involve refinement, via Lewontin's back-and-forth process, of initial impressions of the relevant attributes as these are considered rigorously during the definition of interactors. Once interactors are defined, the nature and scale of the attributes and interactors specifies the scale and defining characteristics of the aggregate within which we count frequencies. Although there can be difficulties in operationalizing some theoretical criteria at certain scales (e.g., interaction of sub-entities and “emergent” properties [Goertzel 1992]), keeping a sharp distinction between theoretical units and empirical entities throughout the theory building process should prevent dilemmas concerning the “level” at which selection is acting. Conflicts between levels or the inability to determine what level on which selection acts in a given situation typically arise because traditional units are used which are frequently inappropriate to a research problem (Hull 1980; Kawata 1987).

## Definitions and Methods

### The Interactor as a Functionally Integrated Biological Entity

Although interactors may be either "individuals" or trait groups, as discussed above, in the present chapter we focus on the former class of units, partially because trait groups have been well discussed in recent literature (i.e., Sober and Wilson 1998), but also because "individuals" form the basis for Dunnell and Wenke's (1980; Dunnell 1978, 1980) argument concerning complex societies. Biological entities are complex open systems, composed of interacting, interdependent parts linked by exchanges of energy, matter and information (Costanza et al. 1993; Odum 1971). Biological entities are “open” because they require energy and matter from the environment external to the entity in order to survive, a process called subsistence regardless of whether the entity in question is a single cell or an interdependent colony of commonsensically-defined organisms. The interaction of the entity and its external environment is effected through the phenotype. Each functionally integrated entity possesses a phenotype, including cells, organisms, and colonies.

Not all biological entities, however, are functionally integrated when viewed in the context of a given problem. For some purposes, many entities are included as sub-parts of more inclusive entities (e.g., cells in organs, social insects in a colony). Some entities, such as a herd of cows, may be readily delineated and observed, but are not cohesive wholes because the herd does not interact with an external environment as a cohesive whole, but as a set of smaller independent entities. Since, in selection models, the fitnesses of entities are defined by the interaction of phenotypes and the external conditions, we are concerned to define units that are functionally cohesive, rather than “herds” of independent entities (Kawata 1987).

Much of the literature on this subject has concentrated on defining “emergent” properties as a key to determining whether an entity at a given scale possesses some degree of functional integration or completeness. Emergence, in this context, means that sub-parts of a particular entity are interacting in such a way as to produce effects not seen by simple examination of sub-parts (Minsky 1986; Kawata 1987). The argument is that entities are functional wholes when they possess functions or attributes that are not simply statistical summations of the behavior of smaller-scale parts. Efforts to create methods to operationalize “emergence” have met with problems, not the least of which is determining when sub-entities are “interacting” sufficiently to produce effects not predictable from smaller-scale entities alone. Theorists grappling with this problem have modeled the problem in terms of non-additivity of fitness components (Lloyd 1988) or attributes (Vrba 1984). Goertzel (1992) reviews these arguments and finds them insufficiently general, preferring a mathematical definition of “non-interaction,” arguing that it is easier to define non-interaction and obtain interactive components by subtraction. To a large degree, however, such methods fall short of the mark.

Attempts to create algorithms for identifying emergent characters are deficient to the degree that they consider quantitative relationships among attributes to be the primary source of information on interaction and therefore emergence. As Sober (1984, 1992) points out, the “emergence” of an attribute at a particular scale is not merely a matter of frequencies but of the biology, ecology and “engineering” aspects of a situation. Phenotypes are produced not by interaction terms in an ANOVA matrix but by structural, organizational and engineering relationships between sub-parts at different scales. In a different context, Minsky (1986) criticized efforts to define “emergent” properties (and holistic approaches to science more generally) because they are unnecessary as solutions to scientific problems. The essential feature of systems that “emergence” tries to capture, Minsky argues, are the structural and functional interaction of parts thereby creating integrated wholes at a larger scale. Phenotypes of entities at a particular scale are created from the structure and functions of entities at smaller scales, as well as the interconnection and organization of entities at smaller scales. Modeling the construction of phenotypes as an organizational problem in engineering allows one, we argue, to recognize integration when and where it happens without resort to vague concepts like "emergence."

In functional biology, it is widely appreciated that different levels of organization possess unique properties that are not characteristics of lower levels. Respiration, for example, is not a property of lungs or of any particular organ, but of the interaction of muscles, the nervous system, and the lungs. Considered in this fashion, so-called “emergent” properties are merely functional consequences of the structural organization of smaller scale parts. Disciplines that study various aspects of organismic physiology, for example, provide powerful evidence that “emergent” properties can be fruitfully studied using the reductionist approach of understanding the construction, behavior, and organization of sub-parts. The physiological approach is applicable generally, though, including situations in which sub-parts are not physically bound into a single localized entity.

The physiology of an entity can be defined, therefore, as the functions of sub-entities, their structure and organization, and the functions that result from interactions between sub-units. Physiology describes the “inner workings” of an entity, and outlines in rigorous terms how the various parts of an entity combine to produce the phenotype and interact with the external environment. In general, entities (e.g., organisms) are functionally differentiated internally and functionally redundant with respect to other entities of the same scale. This means that within functionally integrated entities such as organisms, different sub-parts perform different functions, each of which is necessary to the survival of the whole, and because each sub-part performs only one or a few functions out of the total, sub-parts cannot survive on their own.

The entity (interactor) that interacts with the external environment is thus the larger-scale unit possessing the total range of functions integrated into a cohesively functioning whole. The relationship between interactors, moreover, is one of redundancy, as each possesses the total range of functions necessary for survival and reproduction. Selection occurs between functional units that are internally differentiated but integrated and externally redundant.

In evolutionary biology, the existence of functional differentiation and integration of organisms into larger functional wholes has been one impetus for developing multi-level selection theory (Sober and Wilson 1998; D.S. Wilson 1983; E.O. Wilson 1975). Cultural transmission offers clear possibilities for functional differentiation, visible among most human groups now extant. Concomitant with functional differentiation is the interdependence of individuals with differing functions, a prominent feature of our own society. These features identify many groups of individual humans as cohesive wholes, functioning as integrated units within external environments. It is precisely this feature that defines our concept of “complex societies,” as opposed to functionally undifferentiated “simple societies” (Dunnell 1978, 1989; Dunnell and Wenke 1980). This perspective allows us to see that what anthropologists have traditionally studied in complex societies is the physiology of a larger interactive unit, a point to which we return below.

Because functional integration cannot be reduced to the additivity of fitness or any other simple measure, the physiology of each set of entities required by a particular problem must be understood before interactors can be identified in the real world. Flows of matter, energy, and information within and between entities are patterned, and this patterning serves as the primary source of information on functional integration. Methods for studying patterning, however, must differ between disciplines due to differences in the empirical record studied. In sociocultural anthropology, for example, behavior forms the observable part of the discipline's knowledge base. Linking behaviors together into functionally integrated units will require methods for analyzing repetitive patterns of behavior as part of flows of energy, matter and information. Theories from evolutionary ecology that focus upon feeding behavior and foraging strategies are likely candidates for methods of studying the physiology of human systems. In archaeology, many methods for studying functional patterning, including differentiation and redundancy, are already in place (Dunnell 1995; Fuller 1981), further development requires methods and techniques for extracting functional patterning from the data requirements imposed by the archaeological record itself (Dunnell 1995; Madsen 1997). Table 2 summarizes some of the attributes that might be useful for examining variability in functional integration and differentiation for both artifacts and behavior among human populations.

Once defined within the context of a specific discipline and its particular empirical record, functional integration yields the situation-specific definition for interactors and the algorithm for their identification. For use in selection models, entities that are identified as interactors must meet additional criteria that concern whether variation, differences in performance, and inheritance characterize the entities in their ecological situation. All three are necessary for selection to occur.

Interactors and Selection Processes

Combining insights from the literature on the "units of selection" problem two general conclusions emerge: (i) interactors need to be defined relative to a problem and (ii) any given problem in a specific empirical situation, variation, fitness, and reproduction are each required for an entity to undergo selection. This yields the following definition (stated as a decision algorithm) for interactor:

(Definition 1)

X is a unit of selection in the evolution of trait T if and only if:

(1) X's vary with respect to T (either quantitatively or qualitatively)
*and*
(2) Different variants of T confer differing probabilities of survival and/or replication on the X's.
*and*
(3) X's transmit their variants of T with better-than-chance fidelity through successive replications.

These are, of course, merely the three requirements Lewontin (1970) outlined for natural selection to operate on any particular entity, and there still is no better articulation of the defining criteria for interactors at any scale. These criteria, however, are quite general, and although they capture the requirements for selection to operate in any given case, they offer scant guidance for measurement and identification of entities that meet this definition in any specific discipline (Dunnell 1995). Articulation of the theoretical definition with entities and problems of interest requires that we refine these criteria further.

Phenotypic Variation

Once interactors are defined with respect to a given problem, through analysis of functional organization, it is critical to measure the variation that exists within an aggregate for an attribute of interest. This seems trivial, but empirical entities vary along an infinite number of dimensions. No two entities are precisely the same, either in morphology or in behavior. Significant variation, however, is variation in the attribute classes defined by the research problem. If interactors do not vary with respect to the attribute classes of interest, there can be no selection for those attributes (Sober 1984). It is impossible to explain using selection, for example, why a set of individuals have a vertebrate *bauplan* if all entities identified as interactors in an aggregate have the same body plan. More generally, traits that have been selected to fixation leave no trace of their evolutionary history in attribute frequencies, and thus are not amenable to selective explanation at the scale chosen.

Should we still suspect that selection is indeed responsible for the evolution of vertebrate interactors, we would need to move the scale of analysis upward (for all three scales of units) until the interactors chosen do vary with respect to this attribute. Perhaps the interactors would now be defined at the scale of what was formerly the aggregate, and the aggregate would be at a still higher scale. We might expect that these larger scale interactors would exhibit variation with respect to body plan. In common-sense terms, we have moved our analysis from “microevolutionary” to “macroevolutionary” terms, and the principal tool for analysis at this larger scale might well be phylogenetic analysis and the "comparative method" (Eldredge and Cracraft 1980). Unit requirements and definitions of selection models are invariant with scale; the appearance of discontinuities of process across scales likely derives from the use of common-sense entities in models. Scale-invariant theory yields not distinct hierarchical levels of process, but fundamental continuity of natural process coupled with variation in the empirical scales at which processes act (Hull 1980; Kawata 1987).

The phenotype is defined as the sum total of the relevant morphological and behavioral attributes possessed by a particular entity (Dobzhansky 1970). Since an exhaustive description of all empirical variation is not required by the model, and is indeed impossible in practice (Sober 1992; Dunnell 1971), phenotypes are descriptions of the states of individuals over attribute classes of interest defined by particular problems:

(Definition 2)

(1) The phenotypes of individuals $X_1..X_n$ over a morphological or behavioral attribute class are the particular variants $T_1..T_n$ of an attribute class possessed by individuals identified as $X_1..X_n$

*and*

(2) The phenotypes of individuals $X_1..X_n$ must vary in some respect with regard to trait T, and this variation may be either continuous or discrete.

While all attribute classes must be empirically sufficient (i.e., measurable) to be useful for selection models, not all attribute classes in anthropological research problems are equally measurable. Osgood (1951) differentiated between artifacts and behavior principally because of the difference in ease of observation; behavior is empirical and readily observable but is ephemeral and constantly changing. Observations of behavior, therefore, are not replicable and require special methods for observation and recording if required by a particular research problem. Artifacts or material culture are time-transgressive and typically persist for much longer than behavior, offering the potential for replicated observation and study.

Differential Fitness of Interactors

Each individual that acts as a functional entity must interact with its external environment for subsistence. Interaction of an individual with the external environment is mediated by its phenotype. The efficacy and efficiency of an individual’s encounters with the external world are likely to be different than that of other individuals within the same aggregate. Different variants of a phenotypic attribute may contribute to different degrees of success driving interaction with the environment; furthermore, the advantage or disadvantage of a particular phenotypic variant is relative to other variants present in an aggregate. Variants that are disadvantageous in one aggregate may be relatively advantageous in the company of other variants in another aggregate. Finally, phenotypic attributes do not have fitnesses in and of themselves. Fitness is a function of the interaction of a whole interactor with the environment, not individual traits. The relative value of particular traits are “averaged” into the fitness of the whole phenotype. The “fitness” of phenotypic traits, often referred to in the literature on population genetics, is simply a shorthand way of referring to the average fitness of *individuals* possessing a given trait in a given population.

Variation in fitness according to variation in the local external environment is the point of Sober’s (1984) requirement that aggregates of interactors be chosen such that they are all subject to similar causal forces. In practice, this requirement derives from the fact that fitness is caused by engineering and organizational factors *within* particular ecological contexts, the nature of which have rarely been the subject of analysis in evolutionary biology (Wade and Kalisz 1990; Endler 1986). For example, ceramic tempers may vary in their response to firing regime.[4] Some may perform well in high-temperature regimes, others poorly. Whether this engineering difference is translated into fitness differences for individuals (and what direction the fitness differential takes) is dependent upon conditions faced by a local group of individuals. In an environment with plentiful firewood, for example, high-temperature firing regimes may be common, and tempers that fare poorly in such a regime may contribute a low fitness component to individuals holding such variants. In other environments, performances and consequent fitness contributions may easily be reversed. In addition, some traits have performance values that vary depending upon the frequency of the trait in the local population, or upon the density of conspecifics locally (Maynard Smith 1982, 1989). Frequency- and density-dependent fitnesses depend even more critically upon context than the ceramic example discussed above.

[4] The example (greatly simplified) is derived from the work of Feathers (1990, 1993) on ceramic engineering and evolution in the Southeast.

In this way, it is easy to appreciate why the definition of aggregates (as above) is critical. Aggregates are the group of interactors within which performance values are translated into fitnesses by interaction with the physical environment and local conspecifics. For problems in which fitness of individuals (with reference to a given trait) is not envisioned to be frequency- or density- dependent, the relevant aggregate would be defined on the basis of physical or biotic environmental factors that determine fitness. When frequency- or density-dependence is also important, both the spatial distribution of interactors and the physical/biotic environments are significant for defining aggregates. Until aggregates are defined, translation of performance values into estimates of fitness is not possible. Once defined, many methods exist for estimating fitness differentials.

Methods for determining whether fitness differentials exist between individuals in a group are numerous, and are summarized in a book-length treatment by Endler (1986). There are, however, two broad categories of methods: those methods that attempt to deduce potential fitness differentials from the structural and engineering aspects of individuals (Gould and Lewontin 1979; Feathers 1990; Vermeij 1987), and methods which examine fitness a posteriori via frequency characteristics within populations. The former attempt to use information about the ecology and organization of individuals to predict likely differences in fitness in a population. The latter methods attempt to document fitness differentials by examining the results of selection and other processes, and focus upon the frequency structure and history of a population to the exclusion of structural and functional information. These methods are, of course, complementary and each can be implemented in theory designed for explaining cultural change.

In the case of culturally transmitted information, frequency-based a posteriori methods are not well developed beyond Dunnell's (1978) discussion of the frequency expectations for selective neutrality, since extended by Neiman (1995) and Lipo et al. (1997). Since cultural transmission is not restricted to strict parent-offspring transfers of information (Cavalli-Sforza and Feldman 1981; Boyd and Richerson 1985; Dunnell 1989, 1992; Lipo et al 1997), the rich suite of frequency-based methods developed in evolutionary biology (e.g., Endler 1986) will require substantial revision using mathematical models of transmission (e.g., Boyd and Richerson 1985; Pocklington and Best 1996). The advantage of these methods is that they are useful in all anthropological research: because they involve only frequencies of phenotypic traits, they are easily adaptable to many sources of data, including behavior, genetics or artifactual "hard parts".

Engineering and structural methods are better developed for the cultural case, at least in archaeology (e.g., Hoard and O'Brien this volume; O'Brien et al. 1994; Feathers 1990, 1993, this volume). Feathers (1990, 1993) extensively documented engineering performance and cost differences between alternative ceramic paste compositions in the Central Mississippi River valley, which under postulated environmental conditions probably translated to fitness differences, thereby explaining the gradual replacement of other pastes by shell-tempered ceramics. Although most extant examples of these methods for cultural cases deal with changes in morphological and compositional aspects of the phenotype, there is no reason in principle why spatial organization or behavior cannot be addressed by a priori analyses. All that is required is formal models that link spatial or behavioral attributes to differences in efficiency or performance in subsistence systems, for example. Evolutionary ecology is a rich source of such models, particularly the literature on foraging and feeding strategies in different kinds of environments (see Smith and Winterhalder 1992).

### Heritability of Fitness Variation

The last of Lewontin's three criteria (Definition 1) is that variation in fitness must be heritable through replication. Various workers have interpreted this criterion differently, some requiring that the interactor as a whole act as the reproductive unit, others allowing inheritance to occur through the replication of subunits, generically called *replicators* (Dawkins 1976). A simple example suggests that the first possibility is correct. The family dog is an integrated functional entity, as defined above, because different kinds of cells perform different kinds of functions and because each kind of cell requires the other kinds and the organization of all kinds of cells to survive via the survival of the entire dog. In order to reproduce the whole dog, it is necessary to replicate not only the subunits or cells, but also the relationships between the cells. In sexual reproduction, this is accomplished not by sequential reproduction of each somatic cell but by reproduction of the whole organism by means of specialized cells and decoding mechanisms. The function of these specialized cells is to act as carriers of the information necessary to replicate all cells and their relationships via developmental programs. In such situations, it is “dog” that is the reproductive unit, not genetic material inside the dog or smaller functional units, such as organs. To claim that functionally specialized entities within “dog” are the unit of reproduction misses the point.

Regardless of how information is physically transmitted or decoded to produce the phenotype, if information is replicated piecemeal then the unit of replication may be smaller than the interactor; if replicated for the entire interactor at once, then the interactor itself is the unit of replication. In the case of cultural transmission, the nature of replication is uncertain, both because the potential pathways for information interchange are vastly more complex than most genetic situations and because there is no apparent empirical entity responsible for physically passing information. In cultural transmission, it is likely the case that information is replicated and transmitted continuously, with no “generations” and with new information immediately incorporated into the developmental process (Dunnell 1989).

This, however, is not Lamarckian evolution, because although there is continuous inheritance, inheritance is not preferentially adaptive since there is no *a priori* way for interactors to know what alternative characters will have fitness advantages and bias transmission towards higher frequencies of those characters. Certainly individuals make guesses about what is the best choice of several alternatives, but there is no evidence that these guesses do anything more than generate variation upon which selection is later able to act. In fact, as Boyd and Richerson (1985) develop in their models of transmission bias, decisions about transmission and adoption of variants are likely made not through direct evaluation of alternatives but through shorthand rules that are themselves transmitted culturally. Such shorthand guidelines will, having survived selection pressures for some period of time, be adaptive on average over frequently encountered situations. This is no different than saying that animal behavior is adaptive over the long-term, having been fashioned by selection. Intentionality simply has little to do with creating adaptation, although it has everything to do with creating variation out of which adaptation may be fashioned by selection (Rindos 1984; Dunnell 1989, 1992; O’Brien and Holland 1990).

Fortunately, although there are certainly differences between cultural and genetic transmission that will affect the nature of the mathematical models we create to describe the action of processes such as selection, there are a few conclusions we can reach about heritability from results obtained in archaeology. We need to be able to demonstrate that replication is producing *interactors* that are more similar than chance alone would predict. By definition, for fitness to be heritable, “offspring” (those interactors receiving information) must have a greater than random chance of receiving information on fitter attributes from “parents” (those interactors sending information). Because we can rarely observe transmission in progress, heritability is typically calculated from its phenotypic results (Wilson and Sober 1989).

Heritability in genetic transmission may be calculated by pairwise comparisons between the phenotypic attributes of parents with their offspring. Significant correlation of parents’ phenotypes with offspring indicates that the attributes are heritable to some degree. In fact, heritability can be easily defined and measured as the slope of the regression line between parent and offspring phenotypic trait values or by non-parametric methods if the trait is discrete or dichotomous. This method is readily extendable to cultural transmission systems by making a few simple assumptions. If we assume that individuals receive information on constructing phenotypes from a large number of individuals, and if we assume that transmission between two individuals is more likely as the distance between the individuals decreases, then values for phenotypic attributes should be both spatially and temporally autocorrelated when attributes are heritable. Lack of heritability would result in individuals close in space and in time being no more likely to share phenotypes than with individuals further away along each dimension. This method has been in use in archaeology for most of this century under the rubric of seriation, which essentially measures the spatial and temporal autocorrelation of transmission as new attributes are periodically added to the phenotype by invention or innovation (Dunnell 1970; Lipo et al 1997; Neiman 1995).

Selection can, however, disrupt our ability to use this method for some attribute classes, as selection modifies the frequencies of attributes within groups according to their fitness in local environments. Some similarities are analogous, rather than strictly homologous, and some similarities result from both. The only resolution to this dilemma involves spatial and temporal analysis of transmission, so that continuous transmission, multiple source versions of the simple regression method can be created.

## ORGANIZATIONAL CHANGE AND SCALES OF SELECTION IN U.S. HISTORY

The preceding consideration of units and units construction has clear relevance to the problem of complex societies and their explanation through scales of selection. The availability of scale and transmission system free unit definitions imply that selection may act at a number of empirical scales, wherever the twin requirements of internal functional differentiation and functional integration are met. In fact, Dunnell (1978; Dunnell and Wenke 1980) suggested that what anthropologists view as the "process" of the "origin of complex societies" is actually sets of phenotypic changes associated with reorganization of human systems by inclusion of scales of selection higher than individual organisms. In complex societies, individual human beings are functionally differentiated and integrated into larger wholes with distinctive physiologies that operate with individual humans as sub-parts. This approach to functional organization is a powerful tool in the analysis of contemporary situations, as well as the prehistoric past. The example we take up in the present chapter comes from the economic and social history of the United States, primarily because high-quality data exist on the structure of energy and material flows, as well as occupational specialization.

While it is undeniable that the United States has been from its inception a complex society, it is equally true that the economic and social organization of the U.S. has changed greatly over the past two hundred years. Chandler (1977), in his classic history of American business organization, documents change in the basic economic (i.e., functional) unit from individual- or family-run enterprises, with very little functional differentiation, to the industrial business enterprise with high functional differentiation and hierarchical organization. In the early nineteenth century, functional differentiation (as defined above) occurred within local communities, and most small communities were functionally redundant. As Chandler (1977:245) notes, before the 1830's, nearly all "industrial" production in the U.S. was carried on in small shops or in private homes, with workers recruited from local farm populations as needed allowed to return to farm work when other work was completed. By the later twentieth century, several additional layers of functional organization had been superimposed on this pattern, and at the same time change towards larger scales of functional integration yielded change at all lower levels towards greater functional differentiation and interdependence.

The purpose of our analysis is to examine summary statistical data to determine if organizational transformations have impacted in the way energy, matter, and information have been acquired and moved around over the past 150 years of U.S. history. If organizational changes meet the criteria outlined above for interactors, they provisionally reflect addition of higher scales of selection, or in other words, the formation of interactors composed of larger and larger numbers of smaller scale entities. Increases in the scale of interactors are further posited to occur through changes in the organization of vital functions, rather than being a consequence of simple growth in the population of organisms that make up the interactors. To differentiate simple growth, in as many cases as possible, the data depicted represent per capita amounts or constant (i.e., adjusted) dollar amounts. In other cases where absolute numbers are used, as in the growth of railway mileage (see, for example, Figure 9), we argue following Chandler (1977) that absolute railway mileage is significant in terms of organizational change and does not require transformation.

As we noted, for a set of entities to form an interactor, two functional requirements must be met. First, interactors have sub-units which are functionally differentiated (Table 3). In the case of the United States functional differentiation evolves at several scales, over and above the functional differentiation present in small communities in the early nineteenth century. During the period from 1900 to 1950, for example, there has been a small but marked increase in the diversity of specialized occupations for employed U.S. workers (Figure 5). Simultaneously, there is a substantial decline in the number of farmers. Rural farmers represent the majority of tabulated workers that are functionally independent. The loss of farmers indicates that while proportionally fewer people are able to support themselves independently, more people are becoming clerical workers, managers, service workers, and so on. The latter perform separate but complementary functions within a larger system.

Functional differentiation evolves at several scales, not merely among the smallest-scale entities. Chandler (1977, 1990) documents how firms (functional units analogous to organisms) become functionally specialized, and then as scales of organization shift upward, absorb competitors and finally become internally differentiated themselves. From 1879 to 1948 there has been a several fold increase in the development of specialized industries as measured by capital investment standardized to 1929 constant dollars (Figure 6). Through the late nineteenth century, often thought to be a period of heavy industrial development, investment in manufacturing infrastructure was outpaced by investment in food production. Chandler points out that technological changes, which followed reorganization of most functions into larger scale enterprises, were required for heavy manufacturing. Once technical advances occurred, particularly in metals and metal-forming, other specialized manufacturing firms that construct major parts of the modern phenotype (i.e., chemical, petroleum, and major transportation industries) developed.

The second requirement for entities to be interactors is functional integration (Table 3). That is, functionally differentiated individuals forming part of a larger interactor require functions performed by others in order to survive and reproduce. Individual people do not carry enough of the whole phenotype to perform all of the basic functions. One measure of functional integration is the degree to which food is purchased by individuals rather than self-produced. Between the period 1909 - 1954, there is a substantial increase in the per capita expenditure on purchased foods (Figure 7). The increase in canned and frozen vegetables tell a similar story; people have become more dependent on the functions of other parts of the unit in order to obtain the material, such as food, necessary to survive. At the same time, larger sets of functionally differentiated units (in this case industrial firms) have taken over the task of providing food-producing functions for larger segments of the population.

Another way of looking at functional integration is to examine the history of food production (Figure 8). During the first part of the 20th century, there has been a marked decrease in the number of farmers, yet a dramatic increase in the acreage per farm. Farms are becoming bigger but are being operated by fewer people. In other words, fewer farmers indicates less subsistence farmers but more (and bigger) industrial farms. Similarly, while there has been a decrease in the agricultural labor force, the total labor force in the United States has increased several-fold over the same time period. This trend is can also seen by examining the change in the ratio between agricultural labor and total labor; there are fewer farmers supporting more people. As industrial production methods took over agricultural functions, farmers left the land to join other functional categories. The net effect is that few people, by the middle of the twentieth century, produced or were capable of producing their own food. Similar trends exist in other vital categories: clothing, shelter, and other functions once provided by individuals or local groups are now provided by larger functional units to individual people, who are dependent upon the large systems for subsistence.

Interaction is also a component of functional integration. Sub-entities interact with each other in the process of maintaining integration. This interaction is manifested as highly specialized systems by which matter, energy and information are moved across an entity. These systems are analogous physiologically to the blood stream and nervous system. For example, the growth of the railroad and the use of the automobile is tightly correlated to the development of larger functional entities (Figure 9). Increased speed and range of transportation systems had the effect of increasing the effective size of functionally integrated communities. Community size has increased, in fact, to the point where observers using common-sense definitions of “community” no longer view communities as existing and the history of the twentieth century as being the history of the loss of community life (e.g., Kunstler 1993; Berry 1977, 1993).

Similarly, the movement of energy across large distances became possible during this period, allowing integration of functional units across larger distances. Formerly, when power was provided by human or animal power or by water, wind or steam, energy generation and use was essentially a local function. The locality of energy production and use probably served to keep the scale of functional integration low. The ability to produce energy at one location and distribute it widely, obtained with the development of electrical power generation and distribution, removes this constraint (Figure 10). Concomitant with this development, systems of communication (analogous to the nervous system) have rapidly expanded and are necessary within functionally integrated units larger than can be maintained by verbal communication. Televisions, telephones, and radios show a marked increase throughout the late nineteenth and early twentieth centuries, indicating increased functional integration and increases in the scale that cultural transmission affects groups of people uniformly (Figure 11). The patterned growth in the scale of transmission is particularly evident in the succession of television communication over radio. Television represents not only a new way of transmitting *more* information but a way of transmitting *new kinds* of information that permit further functional integration.

One conclusion that can be drawn from these data is that the functional changes related to the evolution of the United States are not merely a result of the growth of population, the absolute dollar value of economic exchange, or any other simple measure of growth. While population increased at a relatively steady pace over the history of the United States, this growth was regularly preceded and accompanied by major structural transformations in the ability to move matter, energy and information (Figure 12). As Chandler (1977) argues, major changes in the way our society is organized economically (and we would argue socially as well) have come about not through the results of adding more people, more money, and bigger infrastructure, but fundamental changes in the way that functions are organized. In evolutionary terms, small communities that were functionally specialized have been reorganized into larger functional entities over time, with the end result that the scale of interactors in our contemporary situation is quite large, possibly approximating the country as a whole for certain classes of environmental interaction. The conclusion must be that our history is not the history of a continuous process of growth and progressive development, as commonly depicted (e.g., Nevins and Commager 1956), but a history of two basic scales of functional organization with a major evolutionary transformation between them. The nineteenth century and the Industrial Revolution are, for American society, a history of the scale of interactors shifting from something approximating human scales to functional units vastly larger than ourselves.

## CONCLUSIONS

The foregoing has not been an exhaustive study of the problems inherent in creating units to employ within selection models for anthropology; nor is it a complete list of methods for translating a research problem into a set of interactors, attributes, and aggregates, let alone measure their empirical counterparts. Our goal has been to convey the importance of matching formal models of evolutionary processes with discipline-specific systematics which allow the models to be used in the explanation of empirical situations. We have outlined two kinds of units (theoretical and empirical) required for operationalizing selection models, defined three scales at which units are required, and discussed the different functions of each of the six resulting units within selection models. We have described the unit requirements for only one of these units, the interactor, in detail. Lewontin's (1970) criteria for entities to be acted upon by selection served as our starting point, and after discussing the requirement that interactors be functionally integrated entities, we addressed each of Lewontin's criteria with an eye toward outlining a few potential problems with their evaluation in the context of culturally transmitting populations of human beings.

Our choice of the United States as an example of the use of explicit units for recognizing scales of selection was made because good economic and organizational data exist. The example is also a familiar one, and thus potentially is more convincing than one derived from groups that have long been part of the archaeological record. If evolutionary theory is useful for examining ourselves and reaching significant substantive conclusions, it will be useful for anthropology as well as prehistory. Conversely, if evolutionary theory is useful for the past, it has much to say about our present and future situations.

Evolutionary theory, however, cannot be used fruitfully in a heuristic mode. Rigorous models with explicit units are required to make theory do any practical work. Without units, models will only be able to explain particular cases fortuitously, when the ad hoc units used for measurement happen to coincide with the unit requirements of the incomplete model. Because of this, we believe that the next steps in construction of evolutionary theory suitable for explaining human cultural behavior are development of measuring tools for describing organization in ways that are relevant to selection theory. In this respect, we see the work of Fontana and Buss (1994, 1996) on the mathematics of organization at multiple scales as a logical place to begin the construction of a "grammar" for describing organization with the same rigor that archaeologists have traditionally brought to bear on more easily observable phenomena. Progress in the development of evolutionary theory for anthropology, archaeology, or history simply *requires* that evaluation of models in real-world cases be more than randomly successful, and without the proper tools to measure the quantities important to selection theory, our success in applying evolutionary theory to cultural phenomena may be fortuitous at best.

ACKNOWLEDGEMENTS

The authors thank Roger Lohmann for the invitation to participate in NASA's invited symposium at the 1993 annual meeting of the American Anthropological Association, where we presented two related papers which form the basis for this work. Robert Leonard, Lee Lyman, Robert Dunnell, Eric Smith, Stephen Cole, Kim Kornbacher, Sarah Sterling, and Kris Wilhelmsen read the written drafts of one of the conference papers and offered valuable editorial advice for which we are grateful.

**FIGURES**

**TABLES**

**Table 1: Commonsense Units**

# Commonsense Units

### Biology
- Organism
- Population
- Species

### Anthropology
- Civilization
- Community
- Culture
- Population
- Site
- Society
- State

**Table 2: Proposed Methods for Identifying Functional Patterning in Human Groups**

Behavior

| Differentiation | Functional Integration |
|---|---|
| Not everyone performs the same tasks, including food production but also teaching, healing and other necessary work | Individual people rely upon the products of others' work in order to survive |
| Particular tasks are spatially segregated in spaces associated with different groups | Products and services are consumed dominantly within the group, with surpluses not needed for survival moving outside the group |

Artifacts/Manufactures

| Differentiation | Functional Integration |
|---|---|
| Tools and other activity remains are spatially localized | Internally differentiated sets of artifacts and other features will have closer correspondence of artifact styles within the set than with other contemporaneous sets |
| Flows of materials into a set of occupational remains will be spatially heterogeneous, as will conversion of materials into products | Integrated sets of occupational remains will show a wider range of products moving between differentiated sub-units than are moving to points outside the larger-scale unit |

**Table 3: Translating Theoretical Requirements Into Empirical Measurements: U.S. Case**

| | | |
|---|---|---|
| **Theoretical Requirement** | Sub-units are functionally differentiated (perform different functions) at several scales | The interactor and its sub-units require functions provided by other parts of the larger unit to survive (functional integration) |
| **Examples of Empirically Measurable Variables in the U.S. Case** | Increasing occupational diversity<br><br>Increasing specialization of economy at large scales | Reliance of individuals on processed foods<br><br>Development of larger scale systems for moving energy, matter, and information within large-scale unit |

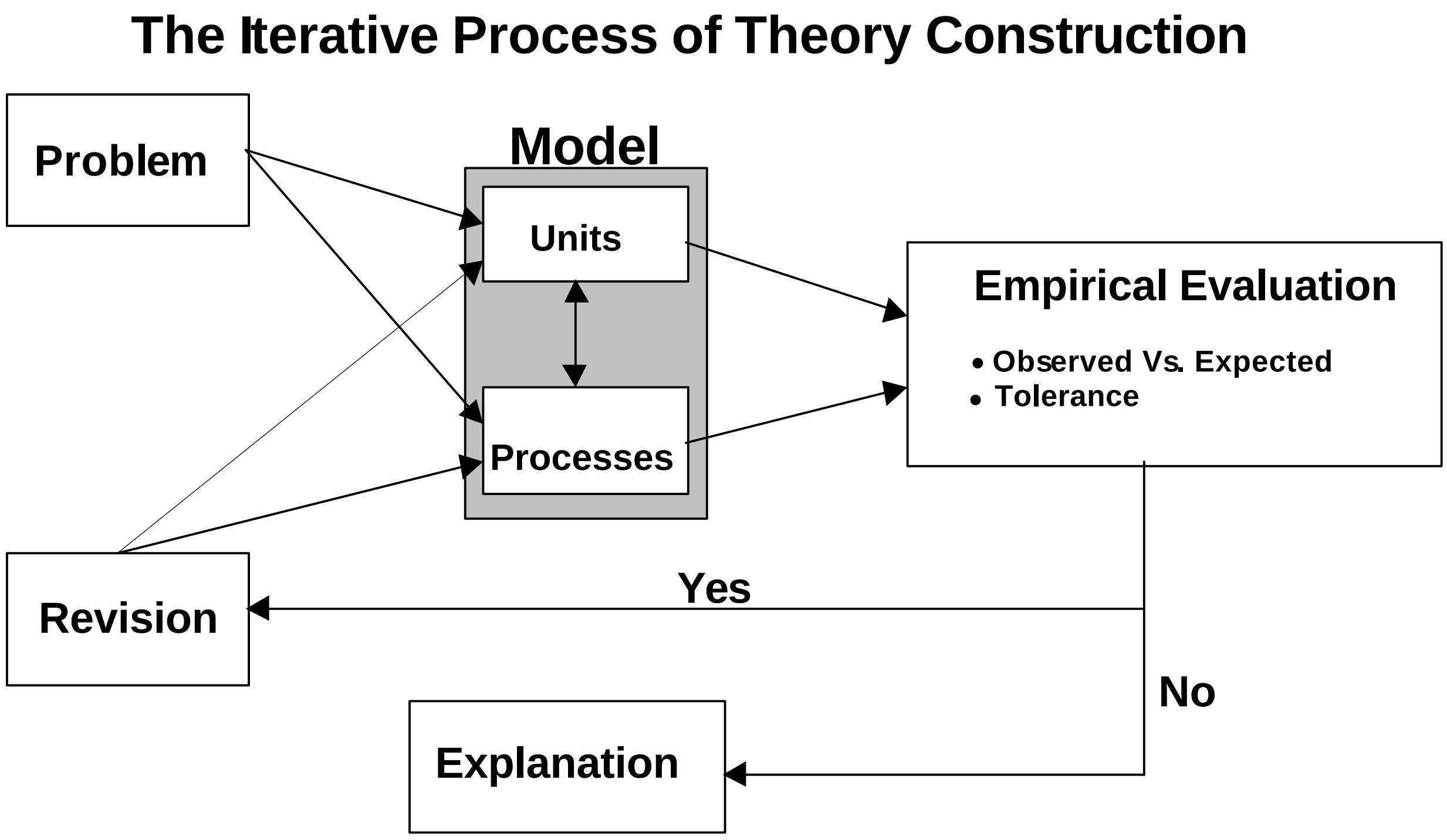

**Figure 1: The Iterative Process of Theory Construction**

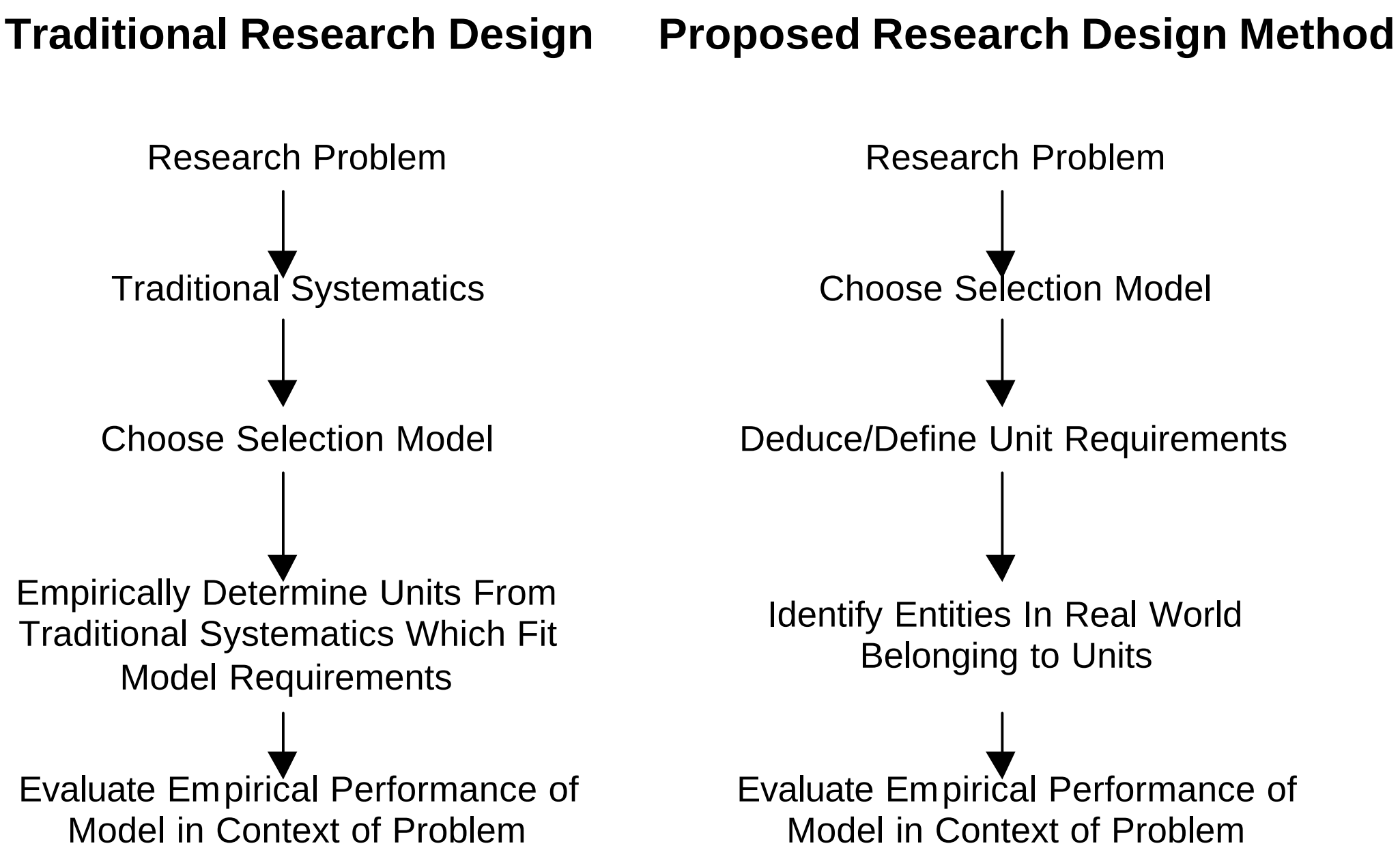

**Figure 2:** **Hull's (1980) Distinction between Uses of Traditional Systematics and Problem Oriented Units**

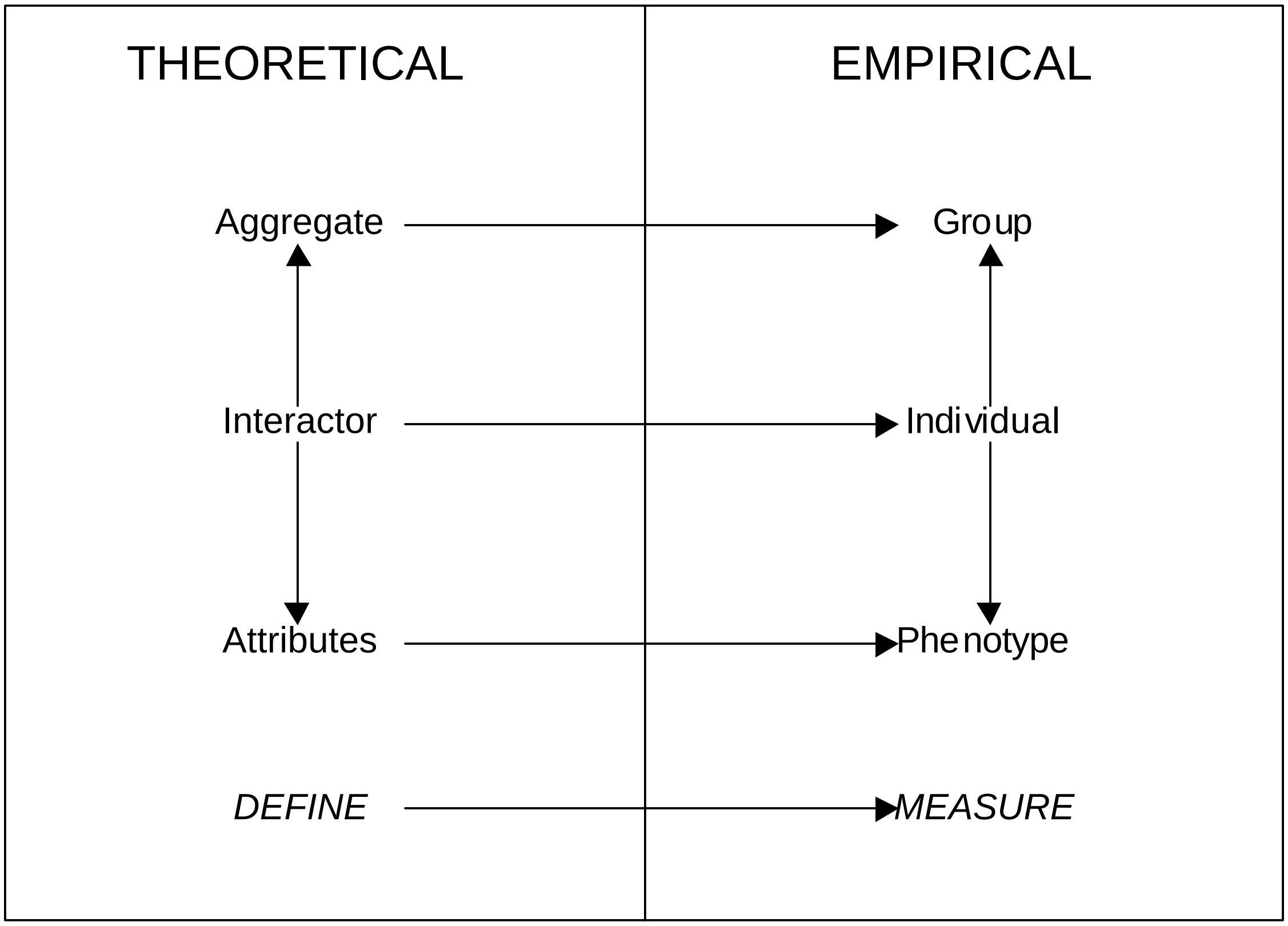

**Figure 3: The Relationship between Theoretical and Empirical Units**

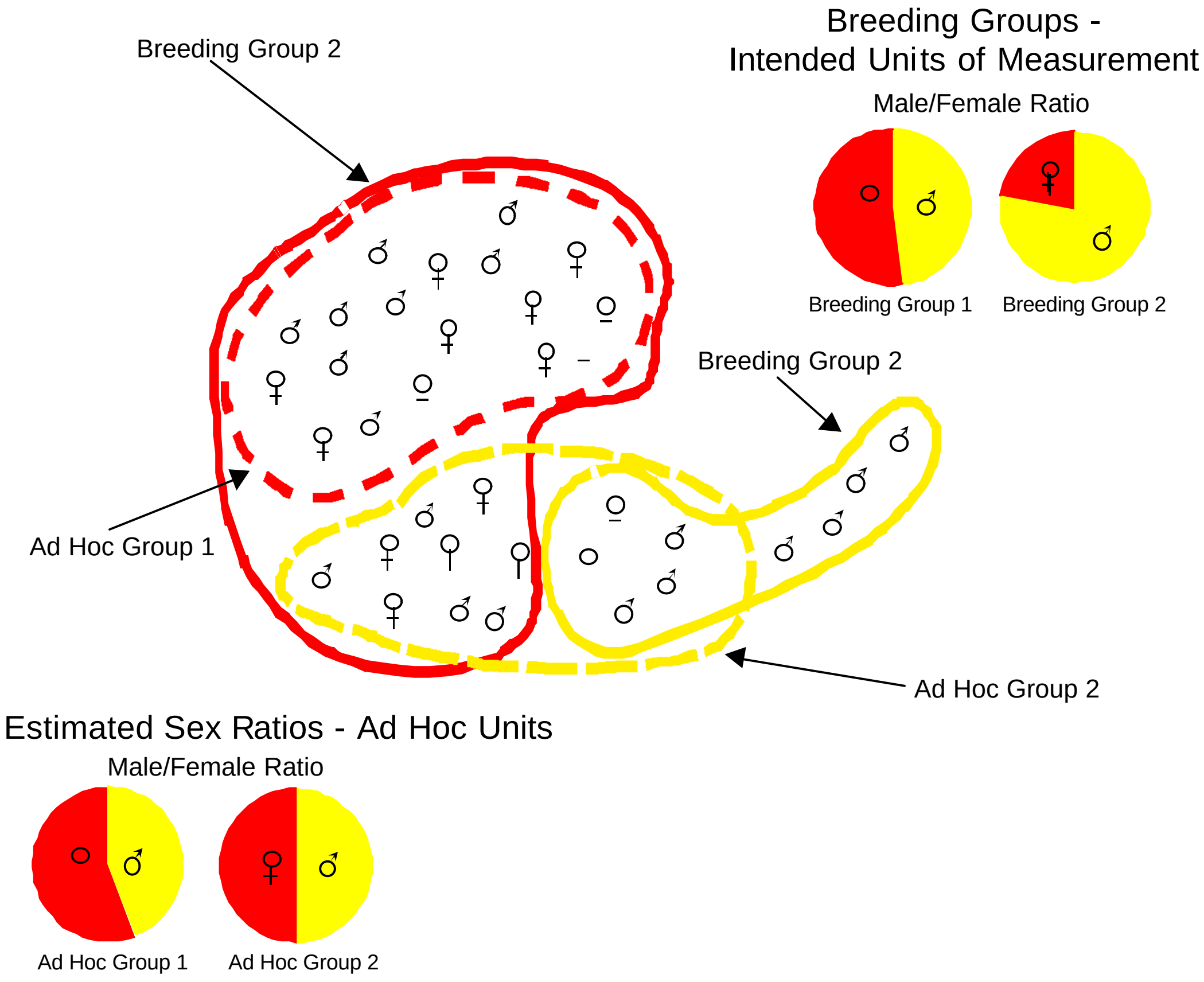

**Figure 4: Effects of Ad Hoc Group Definitions upon Frequencies of Phenotypes**

Proportion of Employed Workers by Occupation Class, 1900-1950

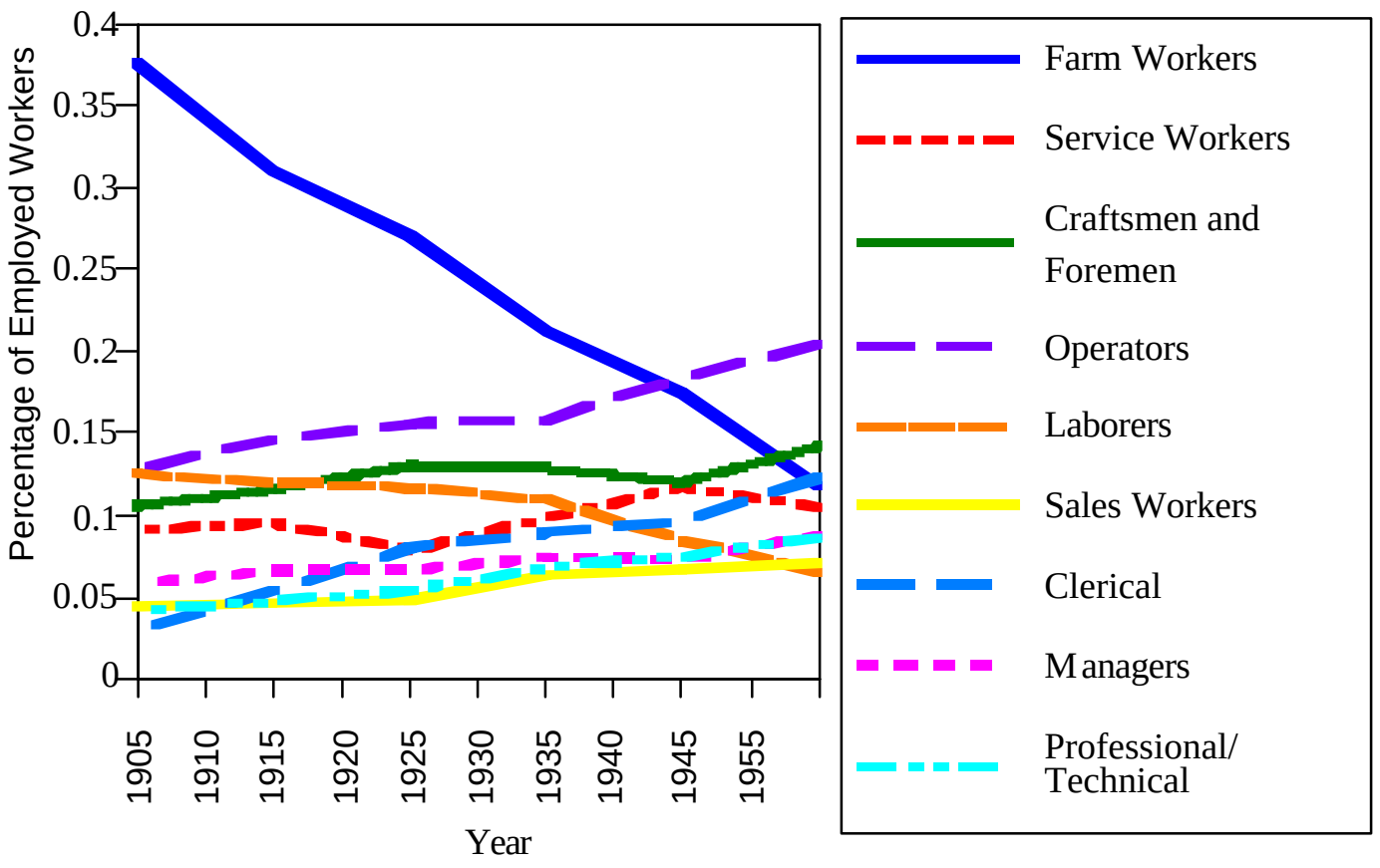

Souce: U.S. Census Bureau, Historical Statistics (1993)

Index of Functional Diversity in the United States, 1900-1950

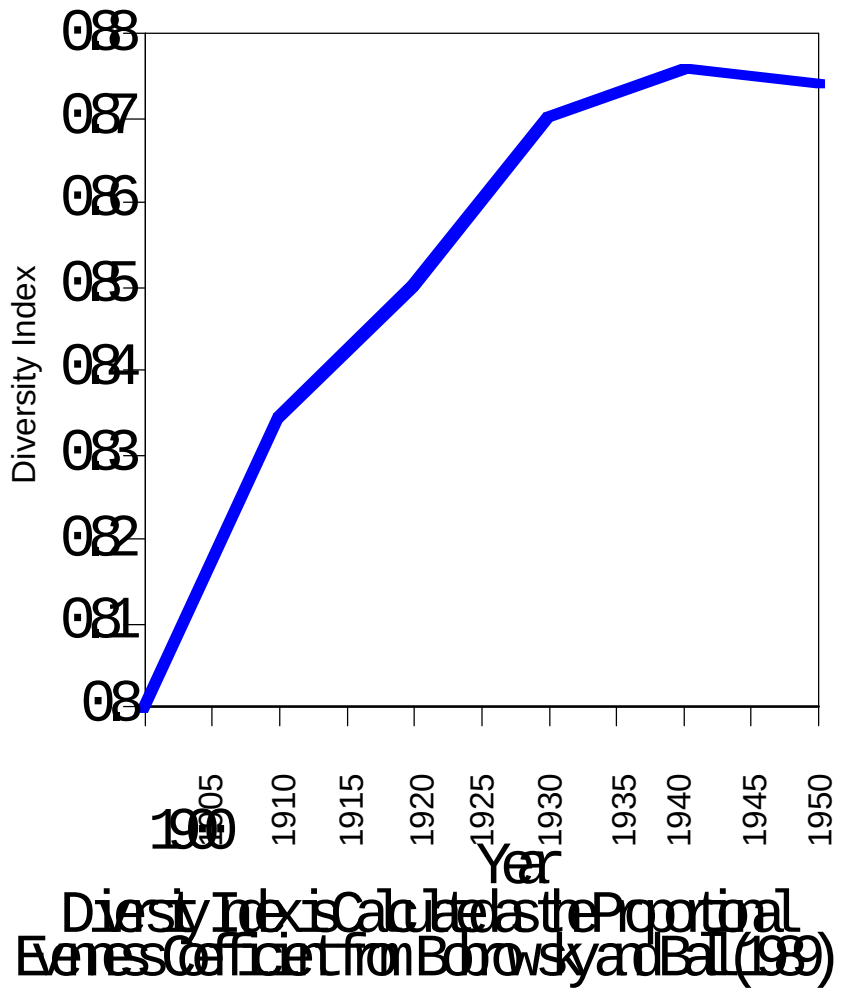

**Figure 5: Growth of Functional Differentiation: Employment**

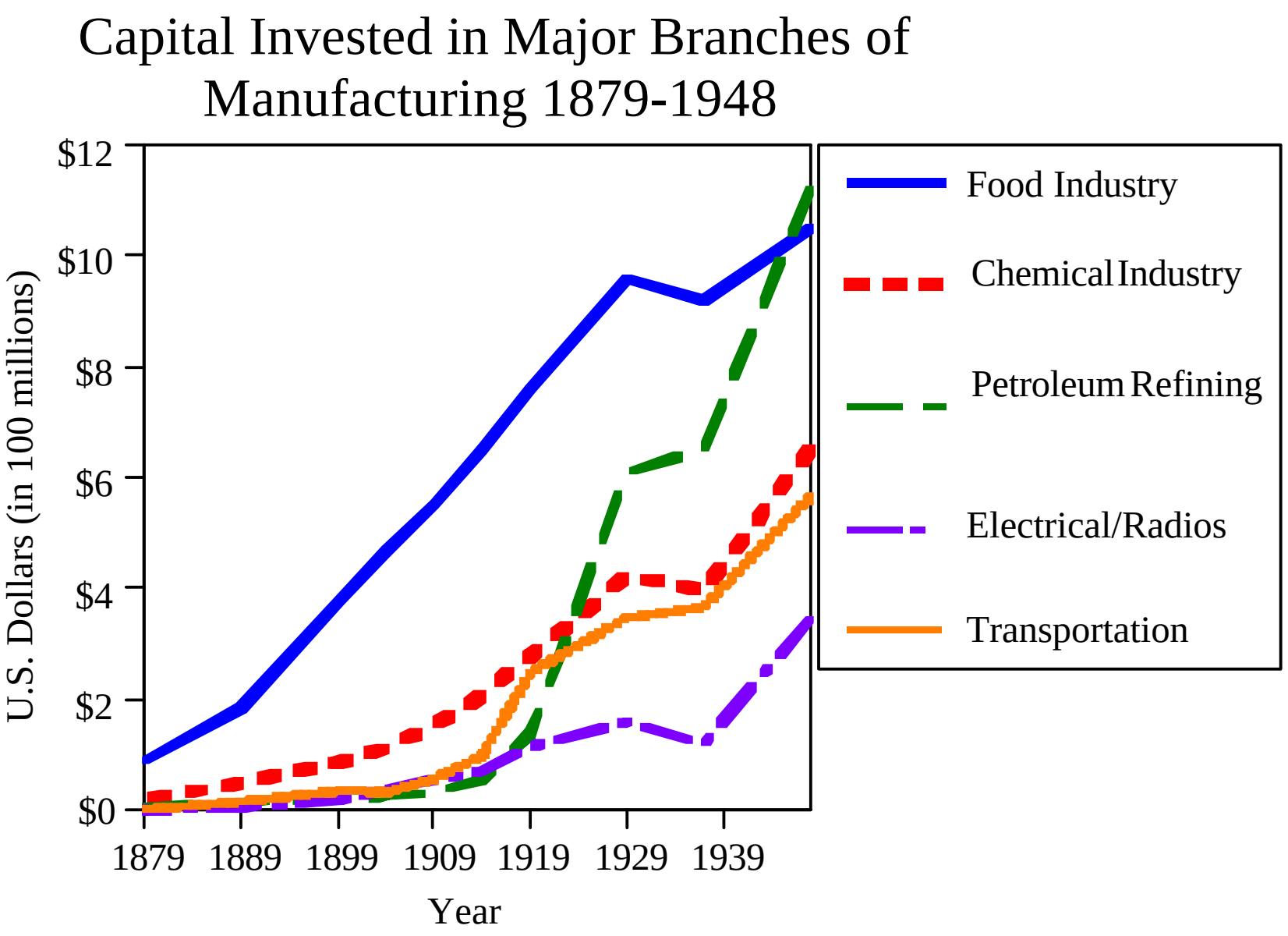

Source: US Census Bureau, Historical Statistics of the United States (1993)

**Figure 6: Growth of Functional Differentiation: Industry**

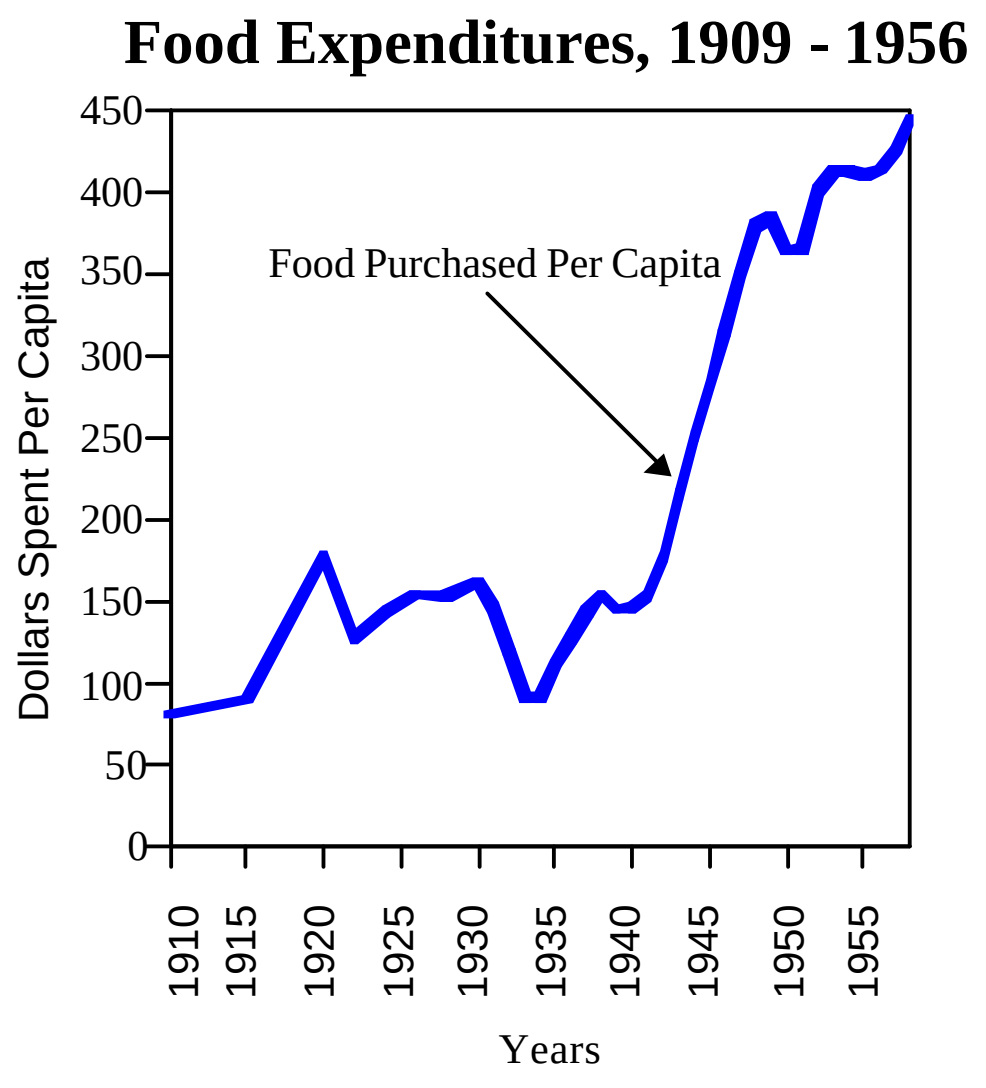

Source: US Census Bureau, Historical Statistics of the United States (1993)

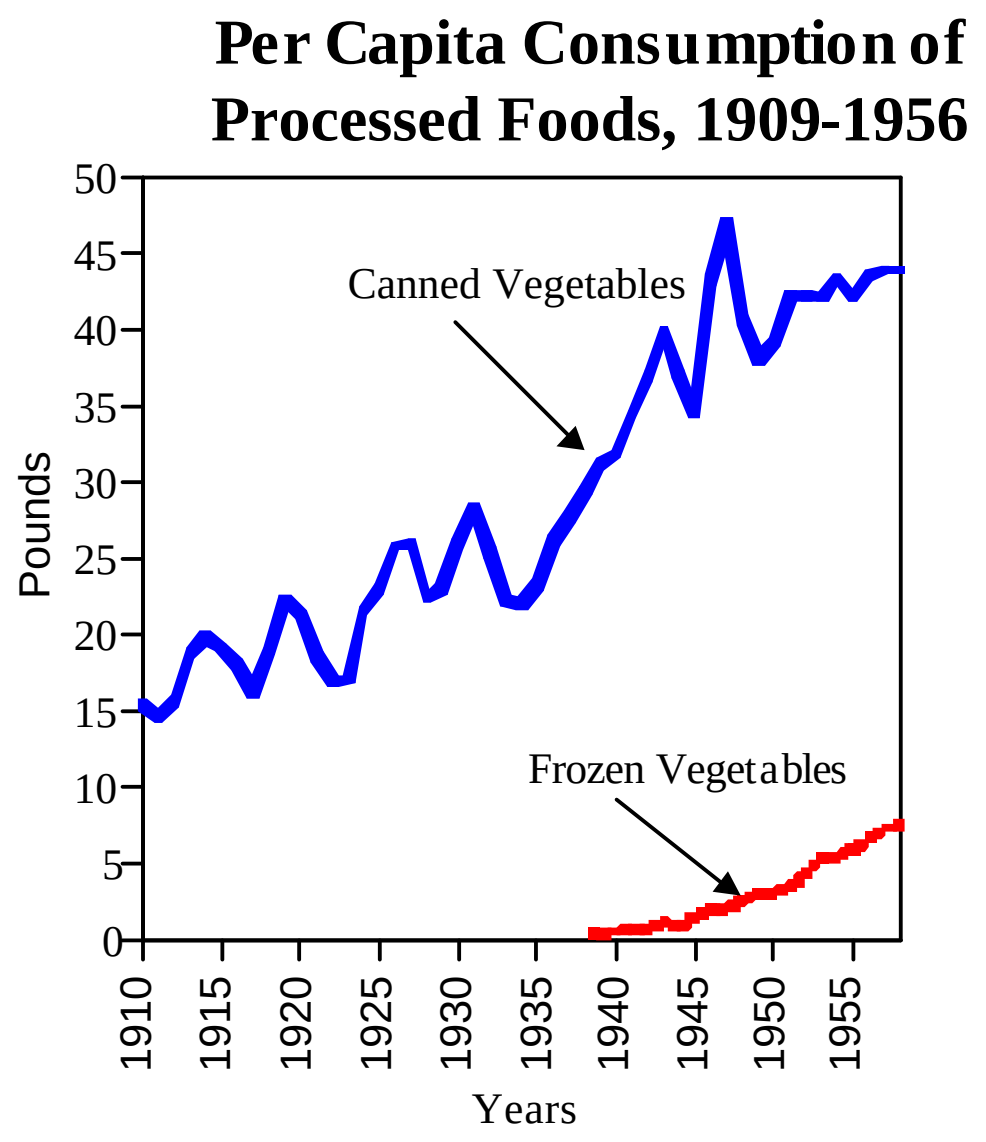

Source: US Census Bureau, Historical Statistics of the US (1993)

**Figure 7: Growth of Functional Integration: Processed Food**

Number of Farms and Acres of Land in Farms, U.S. 1900-1987

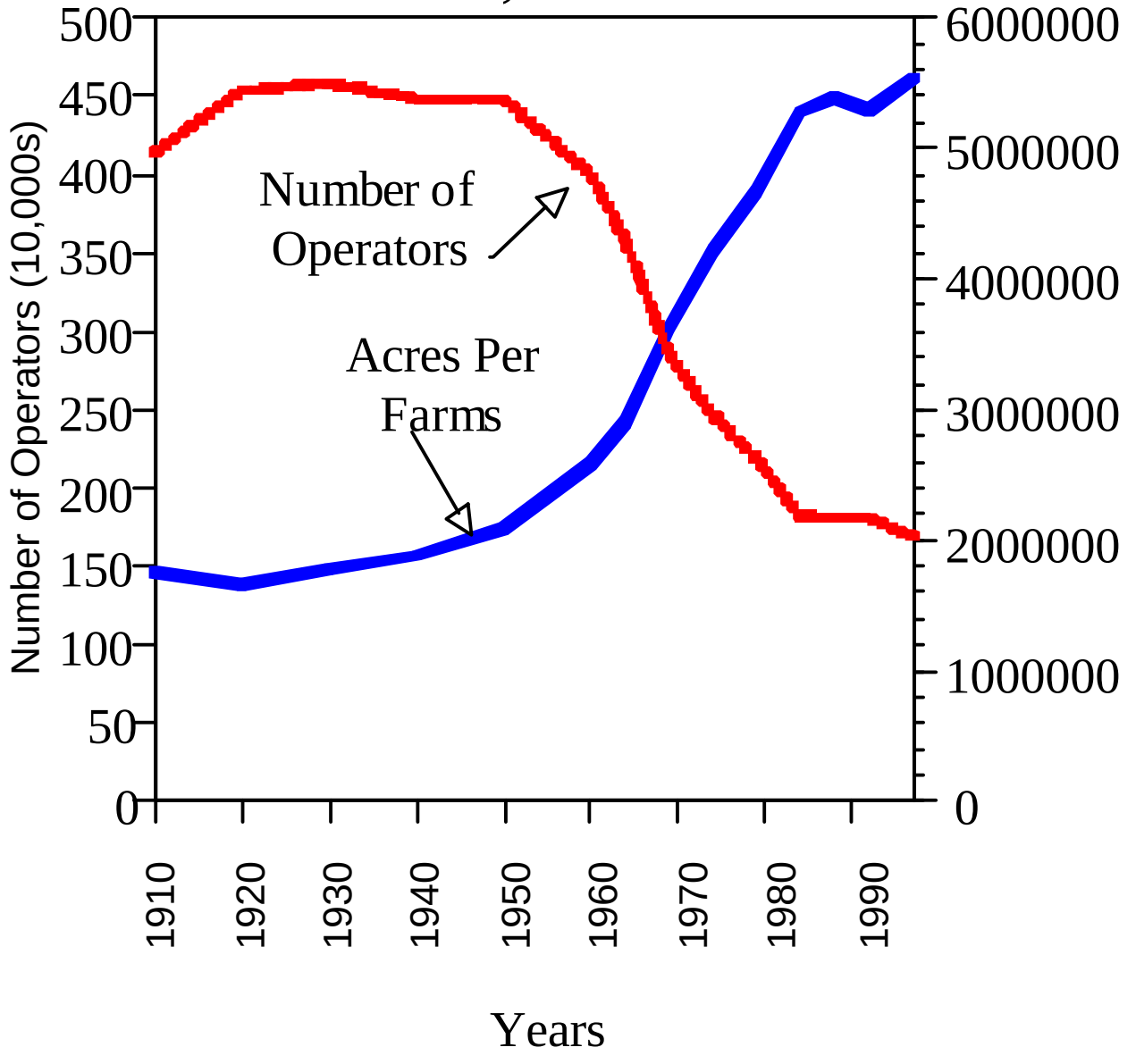

Source: Census of Agriculture for U.S., Bureau of the Census, Washington, D.C. in Demissie (1990:2)

Reductions In Agricultural Work Force, 1910 - 1970

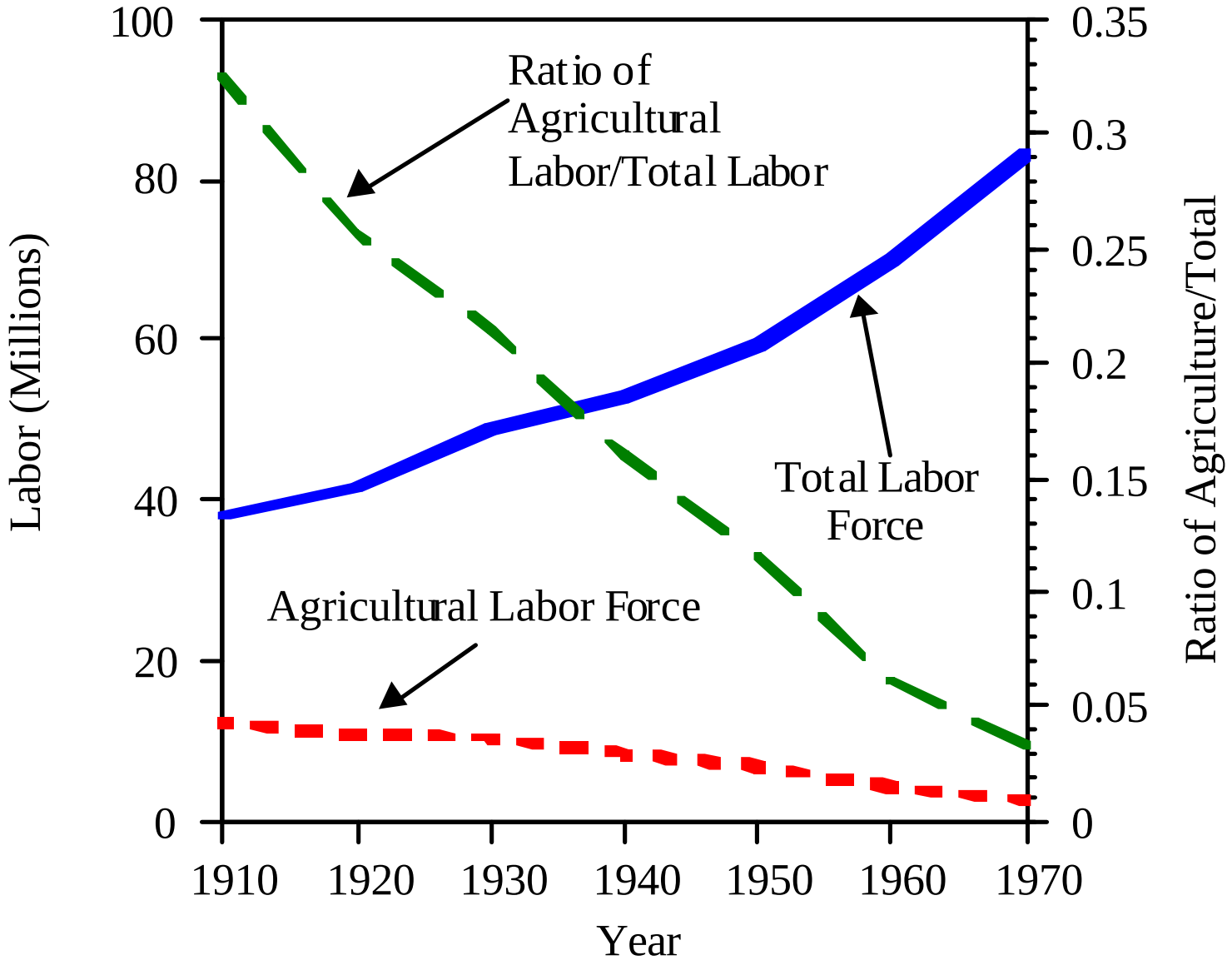

Source: U.S. Bureau of the Census, Historical Statistics of the U.S. (1993)

**Figure 8: Functional Integration: Food Production**

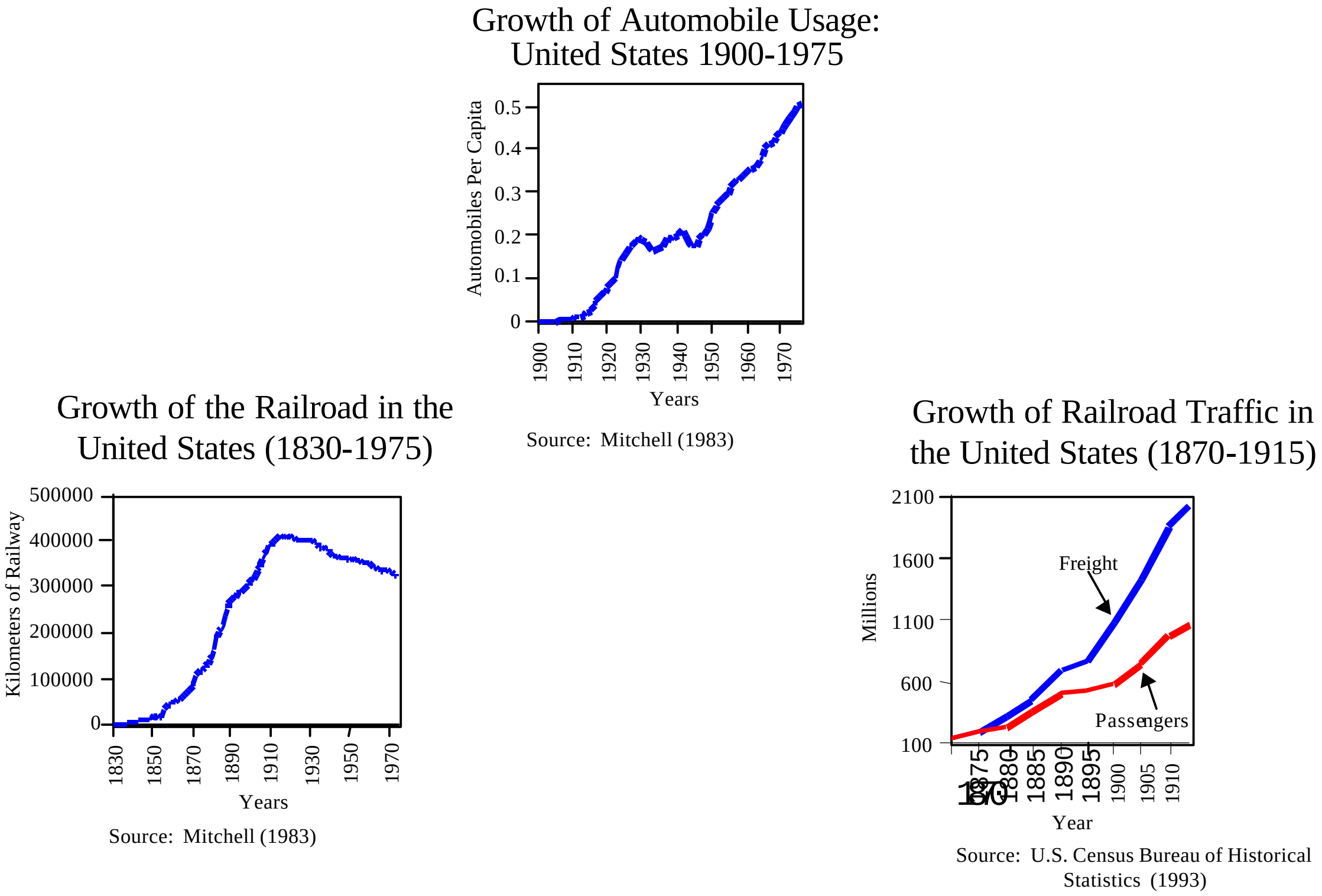

**Figure 9: Functional Interaction: Transportation**

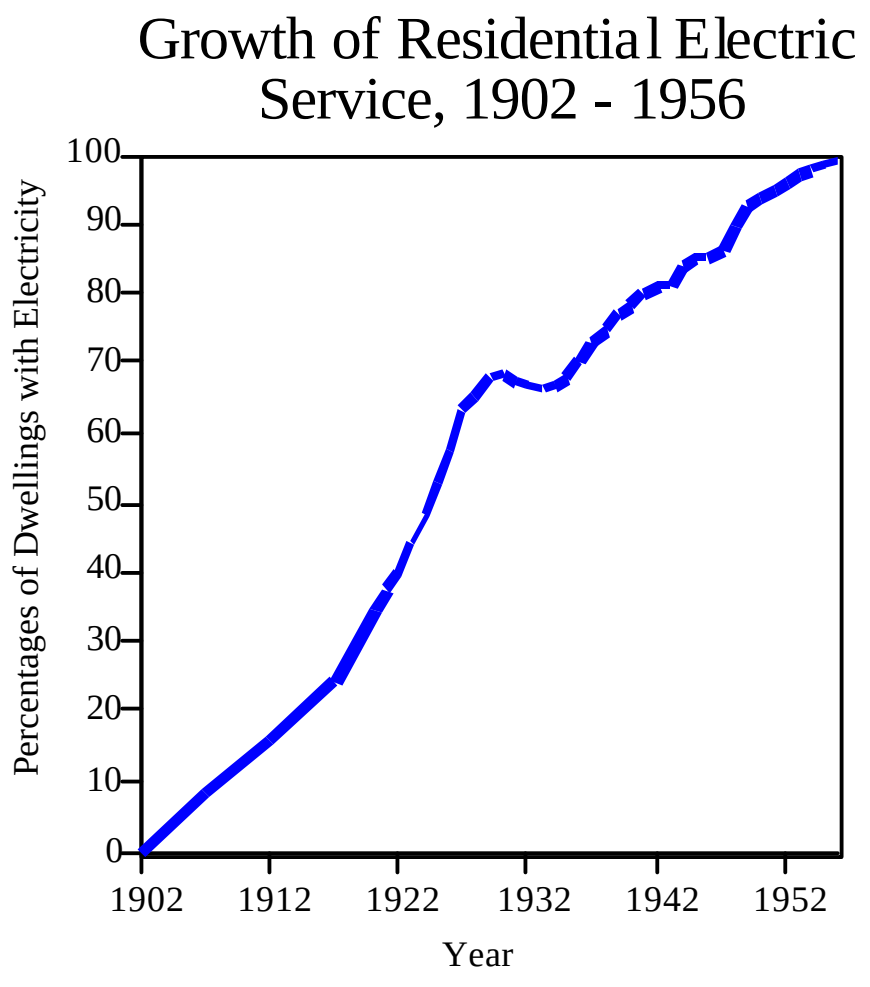

Source: U.S. Census Bureau, Historical Statistics of the United States (1993)

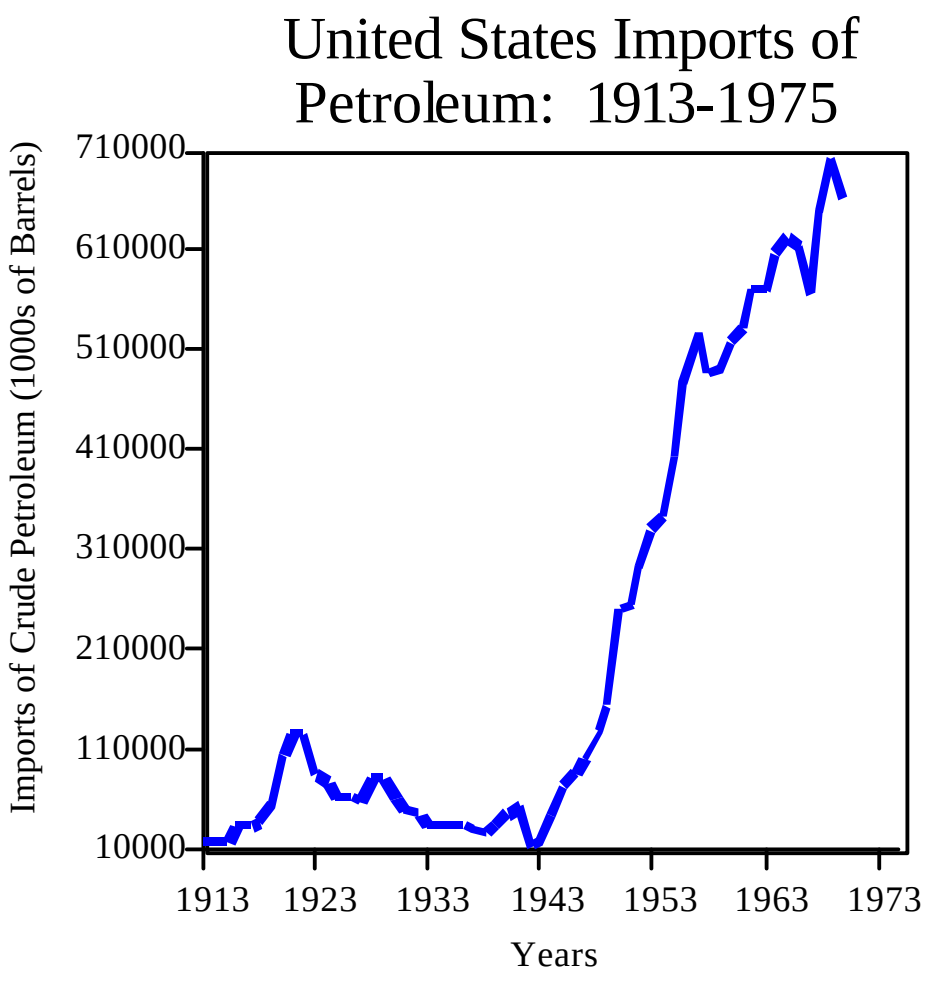

Source: Mitchell (1983)

**Figure 10: Functional Interaction: Energy**

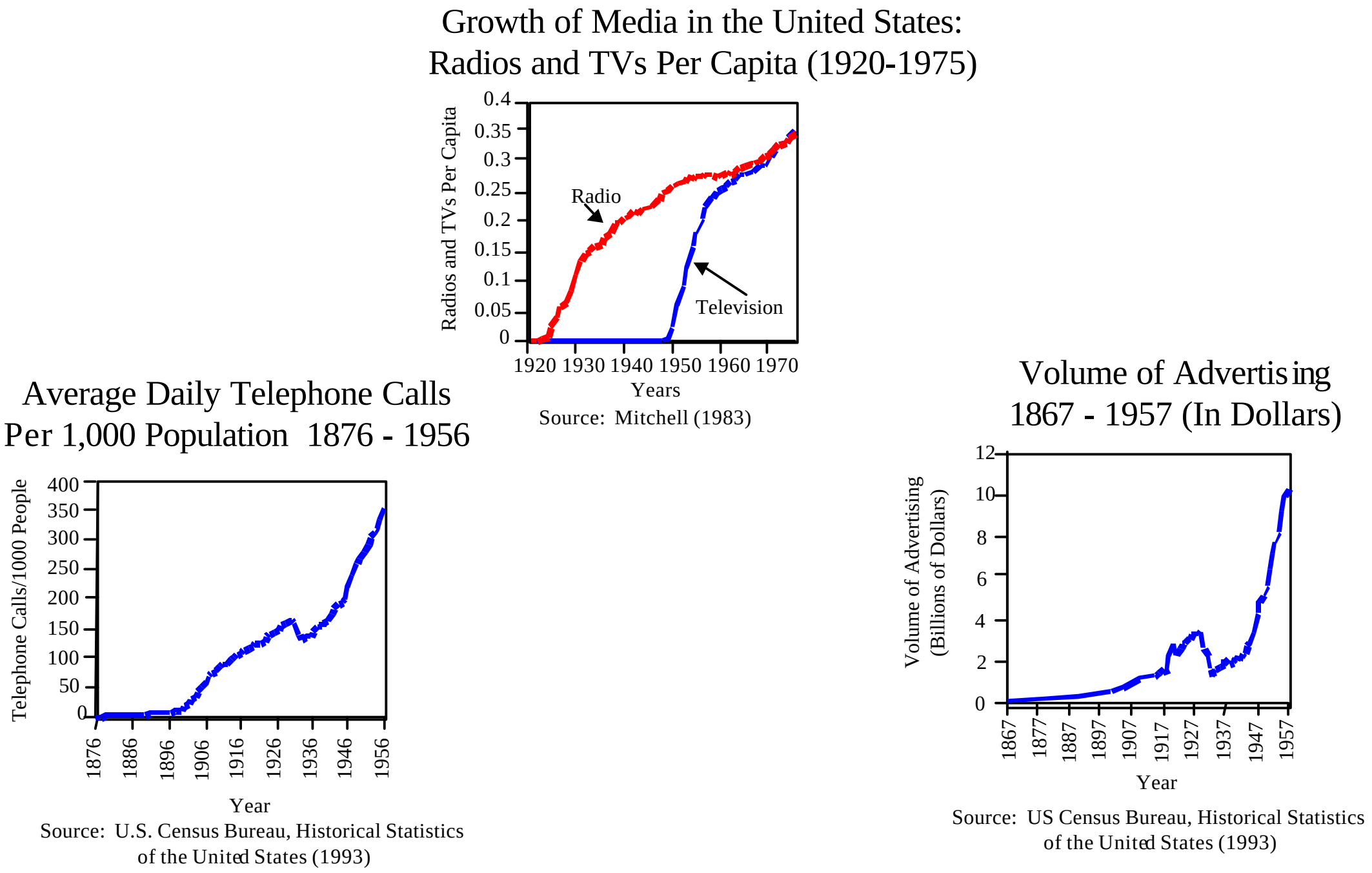

**Figure 11: Functional Interaction: Information**

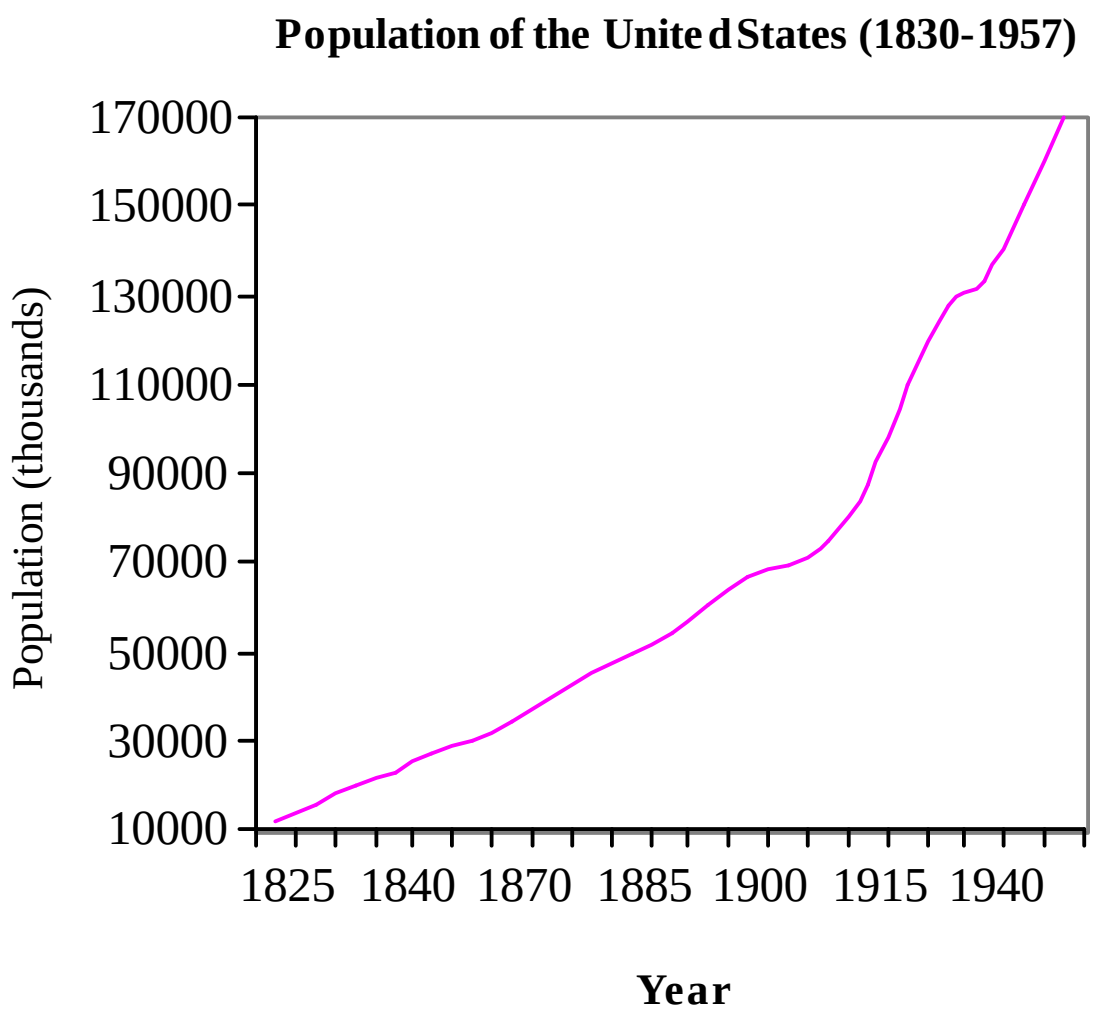

**Figure 12: Population of the United States (1830-1957)**